\documentclass[twocolumn,epjc3]{svjour3}  
\usepackage{amsmath,amssymb,amsfonts}
\usepackage{graphicx,graphics}
\usepackage{graphics}
\usepackage{url}
\usepackage{slashed}
\usepackage{color}

\newcommand{\ppbar}{p--\ensuremath{\bar{\rm p}}}
\newcommand{\PbPb}{Pb--Pb}
\newcommand{\pPb}{p--Pb}
\newcommand{\pp}{p--p}

\journalname{Eur. Phys. J. C}

\begin{document}

\title{Scaling properties of inclusive W$^\pm$ production at hadron colliders}

\author{Fran\c{c}ois Arleo$^{1,}$\thanksref{e1} \and \'Emilien Chapon$^{1,}$\thanksref{e2} \and Hannu Paukkunen$^{2,3,4,}$\thanksref{e3} }

\thankstext{e1}{e-mail: francois.arleo@cern.ch}
\thankstext{e2}{e-mail: emilien.chapon@cern.ch}
\thankstext{e3}{e-mail: hannu.paukkunen@jyu.fi}

\institute{Laboratoire Leprince-Ringuet (LLR), \'Ecole polytechnique, CNRS/IN2P3 91128 Palaiseau, France \and
Department of Physics, University of Jyvaskyla, P.O. Box 35, FI-40014 University of Jyvaskyla, Finland \and
Helsinki Institute of Physics, P.O. Box 64, FI-00014 University of Helsinki, Finland \and
Departamento de F\'\i sica de Part\'\i culas and IGFAE, Universidade de Santiago de Compostela, E-15782 Galicia, Spain
}

\maketitle

\abstract{
We consider the hadroproduction of W gauge bosons in their leptonic decay mode. Starting from the leading-order expressions, we show that by defining a suitable scaling variable the centre-of-mass dependence of the cross sections at the LHC energies can be essentially described by a simple power law. The scaling exponent is directly linked to the small-$x$ behaviour of parton distribution functions (PDF) which, at the high virtualities involved in W production, is largely dictated by QCD  evolution equations. This entails a particularly simple scaling law for the lepton charge asymmetry and also predicts that measurements in different collision systems (p-p, p-$\overline{\rm p}$, p-Pb Pb-Pb) are straightforwardly related. The expectations are compared with the existing data and a very good overall agreement is observed. It is shown that the PDF uncertainty in certain cross-section ratios between nearby centre-of-mass energies can be significantly reduced by
taking the ratios at fixed value of scaling variable instead of fixed rapidity.

}

\section{Introduction}

The production of W gauge bosons in hadronic collisions is a process which is sensitive to practically all aspects of Standard Model, from electro-weak couplings to QCD dynamics and the non-perturbative parton content of the hadrons \cite{Mangano:2015ejw}. One of the most precisely measured observables at hadron colliders is the rapidity ($y$) dependence of the lepton charge asymmetry, ${\mathcal C}_\ell$, 
\begin{equation}
\label{eq:CWdef}
{\mathcal C}_\ell(y) \equiv \frac{d\sigma^{\ell^+}/dy-d\sigma^{\ell^-}/dy}{d\sigma^{\ell^+}/dy+d\sigma^{\ell^-}/dy},
\end{equation}
where the charged lepton ($\ell = e, \mu$) originates from the leptonic decay of the W boson. This observable is a useful probe of proton parton distribution functions (PDFs), in particular, to disentangle the flavour dependence \cite{Berger:1988tu,Martin:1988aj} which is not well constrained by the deep inelastic scattering.\footnote{unless a deuterium target, complicated by possible nuclear corrections, is used.} Today, the charge asymmetry has been studied in detail by the CDF~\cite{Abe:1998rv,Acosta:2005ud} and D0~\cite{Abazov:2007pm,Abazov:2008qv,Abazov:2013rja,D0:2014kma} experiments in \ppbar\ collisions at the Tevatron as well as the ATLAS~\cite{Aad:2010yt,Aad:2011yna}, CMS~\cite{Chatrchyan:2013mza,Khachatryan:2016pev}, and LHCb~\cite{Aaij:2014wba,Aaij:2015zlq} experiments in \pp\ collisions at the LHC. While the broad features of the experimental data are well captured by fixed-order perturbative QCD calculations~\cite{Anastasiou:2003ds,Catani:2009sm}, the simultaneous reproduction of the D0 data in bins of different kinematic cuts is known to pose difficulties~\cite{Ball:2010gb,Lai:2010vv}.

The first measurements of W production in \pPb\ collisions have recently appeared~\cite{Khachatryan:2015hha,Zhu:2015kpa,ATLAS_W_pPb} and various observables seem to  favour the use of EPS09 nuclear PDFs (nPDFs)~\cite{Eskola:2009uj} instead of a naive superposition of free nucleon PDFs (similar conclusion can be expected in the case of other sets of nPDFs~\cite{Hirai:2007sx,Kovarik:2015cma,deFlorian:2011fp}). In addition, these measurements may also help to probe, for the first time, the flavour dependence of nuclear modifications in quark densities~\cite{Khachatryan:2015hha}. The production of W bosons in heavy-ion  collisions is also of paramount importance. Measurements by ATLAS~\cite{Aad:2014bha} and CMS~\cite{Chatrchyan:2012nt} in \PbPb\ collisions have revealed that the production rate approximately scales with the number of binary nucleon-nucleon collisions. This is in sharp contrast to hadronic observables (high-transverse momentum hadrons~\cite{CMS:2012aa,Aad:2015wga,Abelev:2014laa} and jets~\cite{CMS:2012rba,Aad:2014bxa,Adam:2015ewa}) which are strongly suppressed as compared to \pp\ collisions. Thus, the leptons from W decays are valuable ``messengers'' from the initial state of heavy-ion collisions and could also be used to constrain the nPDFs~\cite{Paukkunen:2010qg,Ru:2014yma,Ru:2015pfa}.

In this paper, our main focus is on the centre-of-mass energy ($\sqrt{s}$) systematics of the production cross sections
$d\sigma^{\ell^\pm}/dy$ in hadronic collisions and the consequent scaling properties of the lepton charge asymmetry. First, in Section~\ref{sec:Scalingproperties}, we show how the scaling laws for absolute cross sections and charge asymmetries emerge from the relatively simple leading-order expressions. In Section~\ref{sec:ScalingvsNLO calculation} we then contrast these expectations against next-to-leading order (NLO) computations. Section~\ref{Dataandpredictions} presents comparisons with the existing world data from LHC and Tevatron experiments as well as demonstrates how PDF uncertainties in some ratios of W cross sections can be suppressed by carefully chosing the rapidity binning. Finally, we summarize our main findings in Section~\ref{Summary}.

\section{Derivation of the scaling properties}
\label{sec:Scalingproperties}

\subsection{Absolute cross sections}
\label{sec:AbsScalingproperties}

We consider the inclusive production of W bosons in high-energy collisions of two hadrons, ${\rm H}_1$ and ${\rm H}_2$,
followed by the decay of W to a charged lepton and a neutrino,
$$
{\rm H}_1 + {\rm H}_2 \rightarrow {\rm W}^- + {\rm X} \rightarrow \ell^- + \bar{\nu}
+ {\rm X},
$$
$$
{\rm H}_1 + {\rm H}_2 \rightarrow {\rm W}^+ + {\rm X} \rightarrow \ell^+ + \nu
+ {\rm X}.
$$
At leading order, the production cross section double differential in the charged lepton rapidity $y$ and transverse momentum $p_{\rm T}$ reads~\cite{Aurenche:1980tp,Baer:1990qy}, \vspace{0.2cm}
\begin{eqnarray}
\frac{d^2\sigma^{\ell^\pm}(s)}{dydp_{\rm T}} & = & \frac{\pi p_{\rm T}}{24s^2} \left( \frac{\alpha_{\rm em}}{\sin^2\theta_{\rm W}} \right)^2 
\sum _{i,j} \delta_{e_{q_i} + e_{\overline{q}_j}, \pm1} |V_{ij}|^2  \label{eq:LO1}  \\
& & \hspace{-2cm}
\int_{x_2^{\rm min}}^1 
{\rm d}x_2\ \left(x_2 - \frac{p_{\rm T}}{\sqrt{s}}e^{-y} \right)^{-1}  
\frac{(x_1x_2)^{-1}}{\left(x_1x_2s-M_{\rm W}^2\right)^2 + M_{\rm W}^2 \Gamma_{\rm W}^2} 
\nonumber \\
& & \hspace{-2cm}
\left[
\left(\hat t + \hat u \pm \hat t \mp \hat u \right)^2 q_i^{\rm H_1}(x_1,Q^2) \overline{q}^{\rm H_2}_j(x_2,Q^2) + \right. \nonumber \\
& & \hspace{-2cm} 
\left.
\left(\hat t + \hat u \mp \hat t \pm \hat u \right)^2  \overline{q}^{\rm H_1}_j(x_1,Q^2) q_i^{\rm H_2}(x_2,Q^2) 
\right], \nonumber 
\end{eqnarray}
where the symbols $\alpha_{\rm em}$, $\theta_{\rm W}$, and $V_{ij}$ refer to the fine-structure constant, weak-mixing angle, and elements of Cabibbo-Kobayashi-Maskawa matrix, respectively. The mass and width of the W boson are denoted by $M_{\rm W}$ and $\Gamma_{\rm W}$. The lower limit of the $x_2$ integral is given by $x_2^{\rm min} = ({p_{\rm T} e^{-y}})/({\sqrt{s} - p_{\rm T} e^{y}})$ and the momentum argument $x_1=({x_2 p_{\rm T} e^y})/({x_2 \sqrt{s} - p_{\rm T} e^{-y}})$. The Mandelstam variables $\hat t$ and $\hat u$ are 
\begin{equation}
\hat t = -\sqrt{s} p_{\rm T} x_1 e^{-y}, \quad
\hat u = -\sqrt{s} p_{\rm T} x_2 e^{y}. 
\end{equation}
The PDFs are denoted by $q_i^{{\rm H}_k}(x,Q^2)$ (with $Q^2={\cal O}\left(M_W^2\right)$) and the sum runs over all flavours $i,j$ such that the electric charges $e_{q_i}$ of the quarks sum up to $\pm 1$. Since the total width of the W boson is much smaller than its mass, $\Gamma_{\rm W} \ll M_{\rm W}$, we can make use of a delta-function identity ${\epsilon}/({x^2+\epsilon^2}) \rightarrow \pi \delta(x)$, as $\epsilon \rightarrow 0$, to perform the remaining integral in Eq.~(\ref{eq:LO1}). We find \vspace{0.2cm}
\begin{eqnarray}
\frac{d^2\sigma^{\ell^\pm}(s)}{dydp_{\rm T}} & \approx & \frac{\pi^2}{24s} \left( \frac{\alpha_{\rm em}}{\sin^2\theta_{\rm W}} \right)^2 
\frac{1}{M_{\rm W} \Gamma_{\rm W}}
 \label{eq1} \\
& & \frac{p_T}{\sqrt{1-4p_T^2/M_{\rm W}^2}} \sum _{i,j} |V_{ij}|^2 \, \delta_{e_{q_i} + e_{\overline{q}_j}, \pm1}  \nonumber \\
& & \hspace{-0.5cm}
\left\{
\left[1 \mp \sqrt{1-4p_T^2/M_{\rm W}^2}\right]^2 q_i^{\rm H_1}(x_1^+) \overline{q}^{\rm H_2}_j(x_2^+) + \right. \nonumber \\
& & \hspace{-0.19cm}
\left[1 \pm \sqrt{1-4p_T^2/M_{\rm W}^2}\right]^2 q_i^{\rm H_1}(x_1^-) \overline{q}^{\rm H_2}_j(x_2^-) + \nonumber \\
& & \hspace{-0.37cm}
\,\,\,\, \left[1 \pm \sqrt{1-4p_T^2/M_{\rm W}^2}\right]^2 \overline{q}^{\rm H_1}_j(x_1^+) q_i^{\rm H_2}(x_2^+) + \nonumber \\
& & \hspace{-0.17cm}
\left.
\left[1 \mp \sqrt{1-4p_T^2/M_{\rm W}^2}\right]^2 \overline{q}^{\rm H_1}_j(x_1^-) q_i^{\rm H_2}(x_2^-) 
\right\}, \nonumber \\ \nonumber
\end{eqnarray}
where the momentum arguments of the PDFs are \vspace{0.2cm}
\begin{eqnarray}
 x_1^\pm & \equiv & \frac{M_{\rm W}^2 e^y \,\,\,\, }{2p_T\sqrt{s}} \left[1 \mp \sqrt{1-4p_T^2/M_{\rm W}^2}\right], \\
 x_2^\pm & \equiv & \frac{M_{\rm W}^2 e^{-y}}{2p_T\sqrt{s}} \left[1 \pm \sqrt{1-4p_T^2/M_{\rm W}^2}\right]. \label{eq:x} \\ \nonumber
 \end{eqnarray}
Let us first consider a situation with\footnote{For simplicity, $y\gg0$ ($y\ll0$) should be understood as $e^{y}\gg 1$ ($e^{y}\ll 1$) in the remainder of the paper.} $y\gg0$, that is, $x_2^\pm < x_1^\pm$. In terms of a dimensionless variable $\xi_1$ (which coincides with $x_1^\pm$ when $p_T \rightarrow M_{\rm W}/2$),
\begin{equation}
\xi_1 \equiv \frac{M_{\rm W}}{\sqrt{s}}e^{y}, \label{eq:xi1}
\end{equation}
the momentum fractions in Eq.~(\ref{eq:x}) become \vspace{0.2cm}
\begin{eqnarray}
 x_1^\pm & \equiv & \frac{M_{\rm W}}{2p_T} \xi_1 \left[1 \mp \sqrt{1-4p_T^2/M_{\rm W}^2}\right], \label{eq:x1} \\
 x_2^\pm & \equiv & \frac{M_{\rm W}^3}{2p_Ts\xi_1} \left[1 \pm \sqrt{1-4p_T^2/M_{\rm W}^2}\right]. \label{eq:x2}
\end{eqnarray}
At sufficiently small $x$, the sea-quark densities at high $Q^2 \sim M_{\rm W}^2$ should be reasonably well approximated by a power law \cite{Gluck:1998xa}
\begin{equation}
x\overline{q}_i(x,Q^2) \approx x{q}_i(x,Q^2) \approx N_i \ x^{-\alpha(Q^2)}, \label{eq:power}
\end{equation}
where the exponent $\alpha(Q^2)>0$ and the normalizations $N_i$ should both be almost flavour independent. Such a behaviour (though not exactly a power law \cite{Ball:1994du}) is expected by considering the small-$x$ and large $Q^2$ limit (the so-called double logarithmic approximation~\cite{Gribov:1981ac}) of Dokshitzer-Gribov-Lipatov-Altarelli-Parisi parton evolution equations \cite{Dokshitzer:1977sg,Gribov:1972ri,Gribov:1972rt,Altarelli:1977zs} and it is also consistent with the observations in deep inelastic scattering \cite{Adloff:2001rw} with the $Q^2$ dependence of the exponent $\alpha(Q^2)$ being roughly logarithmic. However, in what follows, the ``running'' of $\alpha(Q^2)$ does not directly show up since we will always set $Q^2 = M_{\rm W}^2$. For brevity, we will denote $\alpha \equiv \alpha(Q^2=M_{\rm W}^2)$ from now on. By using the approximation Eq.~(\ref{eq:power}) in Eq.~(\ref{eq1}) and trading the rapidity variable $y$ with $\xi_1$, we find
\begin{equation}
 \frac{d^2\sigma^{\ell^\pm}(s,\xi_1)}{dp_{\rm T}d\xi_1} \approx s^{\alpha} \times f^\pm(\xi_1,p_{\rm T},{\rm H_1},{\rm H_2}), \quad y \gg 0,
  \label{eq:3}
\end{equation}
where $f^\pm(\xi,p_{\rm T},{\rm H_1},{\rm H_2})$ is a function that does not depend explicitly
on $s$ or $y$. 
Since the expression of Eq.~(\ref{eq1}) is peaked at $p_{\rm T} \approx M_{\rm W}/2$ and the $p_{\rm T}$ dependence of the probed momentum fractions in Eqs.~(\ref{eq:x}) is not particularly fierce, the $x$ interval spanned by integration over $p_{\rm T}$ with a typical kinematic cut $p_{\rm T} \gtrsim 20 \, {\rm GeV}$ remains sufficiently narrow such that approximation of Eq.~(\ref{eq:power}) stays valid.
Under these conditions, the scaling law in Eq.(\ref{eq:3}) holds also for $p_{\rm T}$-integrated cross sections,
\begin{equation}
 \frac{d\sigma^{\ell^\pm}(s,\xi_1)}{d\xi_1} \approx s^{\alpha} \times F^\pm(\xi_1,{\rm H_1},{\rm H_2}), \quad y \gg 0,
  \label{eq:scalingxsec}
\end{equation}
where $F^\pm(\xi_1,{\rm H_1},{\rm H_2}) \equiv \int dp_{\rm T} f^\pm(\xi_1,p_{\rm T},{\rm H_1}) \theta(p_{\rm T}-p_{\rm T}^{\rm min})$. In the backward direction with $y \ll 0$, the appropriate scaling variable is
\begin{equation}
\xi_2 \equiv \frac{M_{\rm W}}{\sqrt{s}}e^{-y}, \label{eq:xi2}
\end{equation}
such that
\begin{equation}
 \frac{d\sigma^{\ell^\pm}(s,\xi_2)}{d\xi_2} \approx s^{\alpha} \times G^\pm(\xi_2,{\rm H_1},{\rm H_2}), \quad y \ll 0,
  \label{eq:scalingxsec2}
\end{equation}
where $G^\pm(\xi_2,{\rm H_1},{\rm H_2})$ is a function that does not depend explicitly on $s$ or $y$. 
If ${\rm H_1} = {\rm H_2}$, then $F^\pm(\xi_1,{\rm H_1},{\rm H_2}) = G^\pm(\xi_2,{\rm H_1},{\rm H_2})$.

Here, we emphasize the fact that at fixed $\xi_1$ ($\xi_2$) the $x$ region at which the PDFs of hadron ${\rm H}_1$ (${\rm H}_2$) is sampled becomes approximately independent of $\sqrt{s}$, see Eq.~(\ref{eq:x1}). Going to forward (backward) direction pushes this region to large $x$ where the parameterization dependence of PDFs may be large. As a consequence, one could hope that the PDF uncertainties on cross-sections ratios between two different values of $\sqrt{s}$ would better cancel out if performed at fixed $\xi_{1,2}$ than at fixed rapidity (as has been done e.g. by LHCb collaboration \cite{Aaij:2015zlq}). At small $x$, the probed $x$ regions will be different for two different $\sqrt{s}$, see Eq.~(\ref{eq:x2}), but at large $Q^2$ the $x$ dependence is almost purely dictated by the DGLAP evolution (in our scaling law approximated by a power law) and less prone to PDF uncertainties. We will come back to this later on in Section~\ref{Crosssectionratios}.

\subsection{Charge asymmetries}
Since the $\sqrt{s}$ dependence in Eqs.~(\ref{eq:scalingxsec}) and~(\ref{eq:scalingxsec2}) is completely in the common prefactor $s^{\alpha}$, it follows that the lepton charge asymmetry Eq.~(\ref{eq:CWdef}) should obey a particularly simple scaling law, \vspace{0.2cm}
\begin{eqnarray}
{\mathcal C}_\ell^{{\rm H}_1,{\rm H}_2}(s,\xi_{1}) & \approx & F(\xi_{1},{\rm H_1},{\rm H_2}), \quad y \gg 0, \\
{\mathcal C}_\ell^{{\rm H}_1,{\rm H}_2}(s,\xi_{2}) & \approx & G(\xi_{2},{\rm H_1},{\rm H_2}), \quad y \ll 0, \nonumber 
\end{eqnarray}
where 
\begin{equation}
 F(\xi,{\rm H_1},{\rm H_2})\equiv \frac{F^+(\xi,{\rm H_1},{\rm H_2})-F^-(\xi,{\rm H_1},{\rm H_2})}{F^+(\xi,{\rm H_1},{\rm H_2})+F^-(\xi,{\rm H_1},{\rm H_2})},
\end{equation}
and similarly for $G$. In other words, at fixed $\xi_1$ or $\xi_2$, the charge-asymmetry should be approximately independent of the centre-of-mass energy. In fact, here one can allow the exponent $\alpha$ to depend also on $\sqrt{s}$ and $\xi_{1,2}$ and it is only required that the PDFs are \emph{locally} well approximated by a power law in the relevant region at small-$x$.

Another, and also a bit surprising feature of the charge asymmetry is that at sufficiently large $|y|$ it effectively depends only on the nucleon that is probed at large $x$. This follows from the facts that when $|y|$ is sufficiently large, either $u\overline{d}$ or $d\overline{u}$ partonic process eventually dominates, and that the light-sea-quark distributions are expected to be approximately SU(2) symmetric at small $x$,
\begin{equation}
u(x,Q^2) \approx \overline{u}(x,Q^2) \approx d(x,Q^2) \approx \overline{d}(x,Q^2), \,\,\, x \ll 1, \label{eq:seaquark}
\end{equation}
and thus symmetric with respect to charge conjugation and isospin rotation. For example, one would expect that ${\mathcal C}_\ell^{{{\rm p}},{{\rm p}}}(s,\xi_{1}) \approx {\mathcal C}_\ell^{{{\rm p}},{\overline{\rm p}}}(s,\xi_{1})$ at large $\xi_1$. In the case of nuclei the nPDFs $f_i^{A}(x,Q^2)$ are built from the free nucleon PDFs $f_i^{{\rm proton}}(x,Q^2)$ and nuclear modification factors $R^{{\rm proton},A}_i$ by 
(see e.g. \cite{Eskola:2009uj})
\begin{equation}
f_i^{A}(x,Q^2) = Z f_i^{{\rm proton},A}(x,Q^2) + N f_i^{{\rm neutron},A}(x,Q^2),
\end{equation}
where \vspace{0.2cm}
\begin{eqnarray}
f_i^{{\rm proton},A}(x,Q^2)  & = & R^{{\rm proton},A}_i f_i^{{\rm proton}}(x,Q^2), \\
f_i^{{\rm neutron},A}(x,Q^2) & = & f_{i, u \leftrightarrow d}^{{\rm proton},A}(x,Q^2). 
\end{eqnarray}
At small-$x$ one expects modest shadowing ($R^{{\rm proton},A}_i<1$) which, however, should not significantly alter the scaling exponent $\alpha$ (particularly at high $Q^2 \sim M_{\rm W}^2$ involved here) and, to a good approximation, the effect of shadowing is just a slight overall downward normalization in the absolute cross sections which should largely disappear in the case of charge asymmetry. In other words, we can encapsulate the scaling law for lepton charge asymmetry as \vspace{0.2cm}
\begin{eqnarray}
{\mathcal C}_\ell^{{\rm H}_1,{\rm H}_2}(s,\xi_{1}) & \approx & F(\xi_{1},{\rm H_1}), \quad y\gg0, \nonumber \\
{\mathcal C}_\ell^{{\rm H}_1,{\rm H}_2}(s,\xi_{2}) & \approx & G(\xi_{2},{\rm H_2}), \quad y\ll0,
\label{eq:masterscaling}
\end{eqnarray}
independently of the nature of hadron (nucleon, anti-nucleon, nucleus)  probed at small $x$.

\section{Scaling vs. NLO calculation}
\label{sec:ScalingvsNLO calculation}

\begin{figure}[htb!]
\centering
\includegraphics[width=\linewidth]{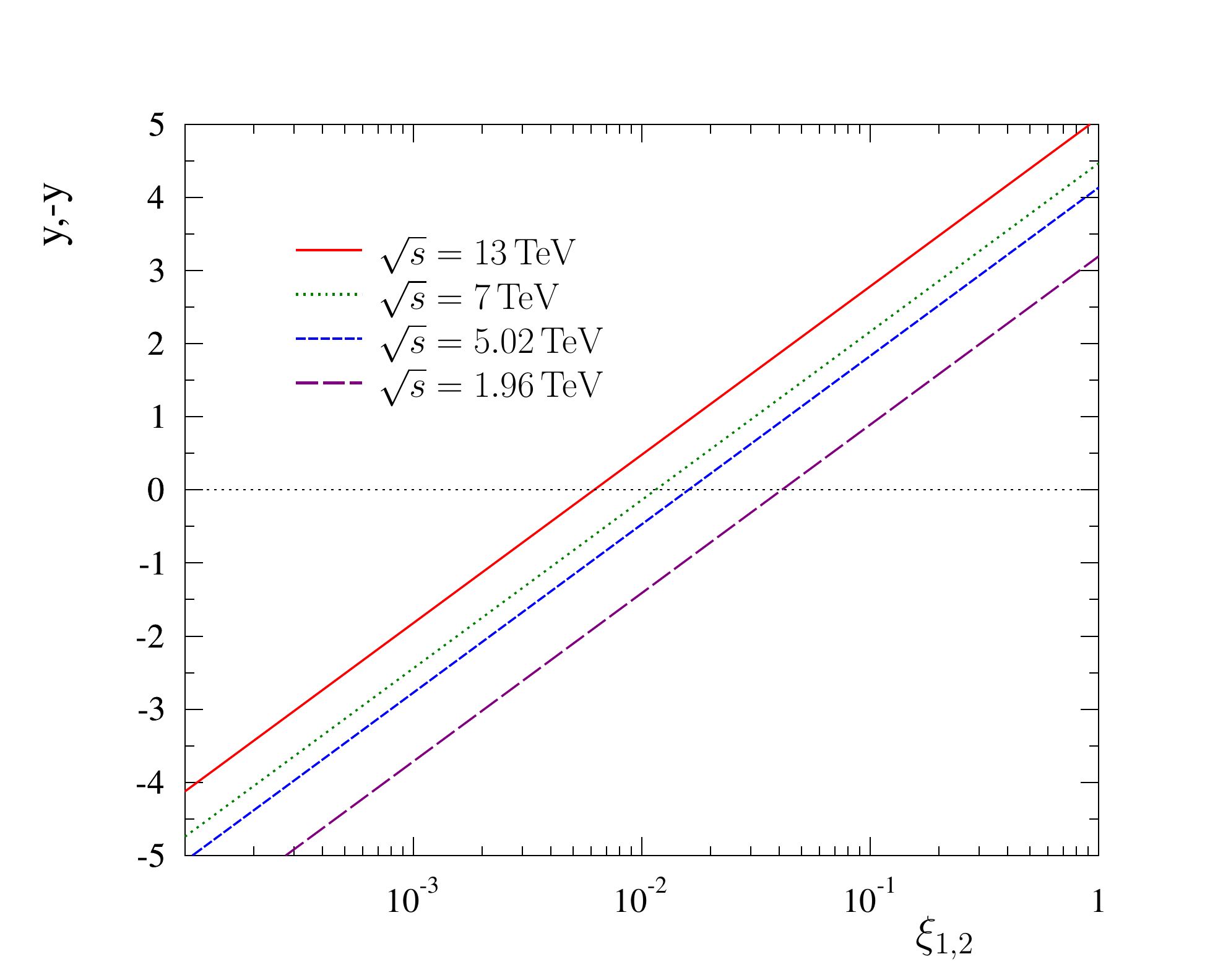}
\caption{Relation of rapidity $y$ and scaling variables $\xi_{1,2}$ for a few values of $\sqrt{s}$.}
\label{fig:xi_vs_y}
\end{figure}

Most of our plots in the rest of the paper will use the scaling variables $\xi_{1,2}$ which are related to rapidity $y$ and centre-of-mass energy $\sqrt{s}$ via Eq.~(\ref{eq:xi1}) and Eq.~(\ref{eq:xi2}). To ease the interpretation in what follows, this dependence is illustrated in Fig.~\ref{fig:xi_vs_y}.

\begin{figure}[htb!]
\centering
\includegraphics[width=\linewidth]{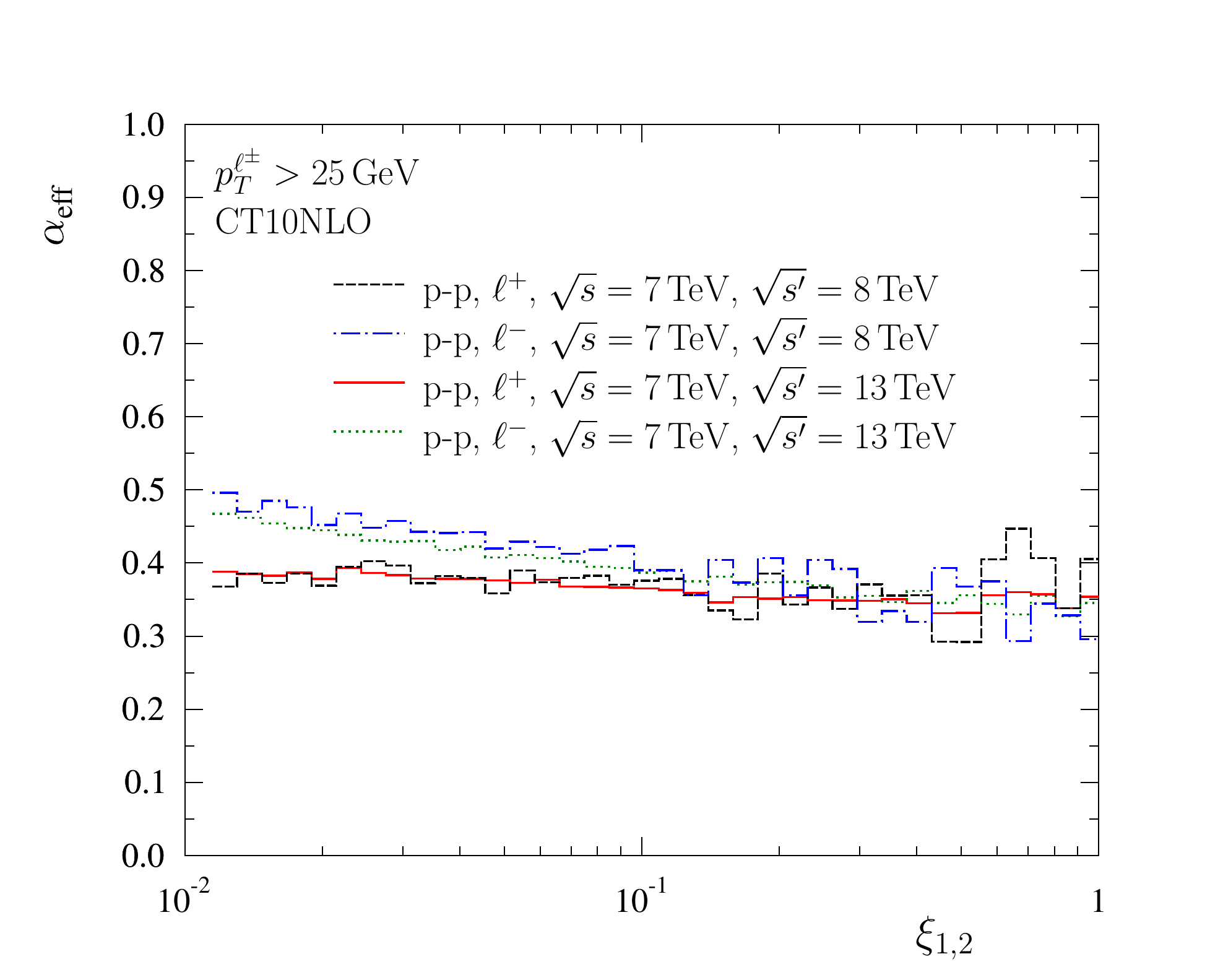}
\includegraphics[width=\linewidth]{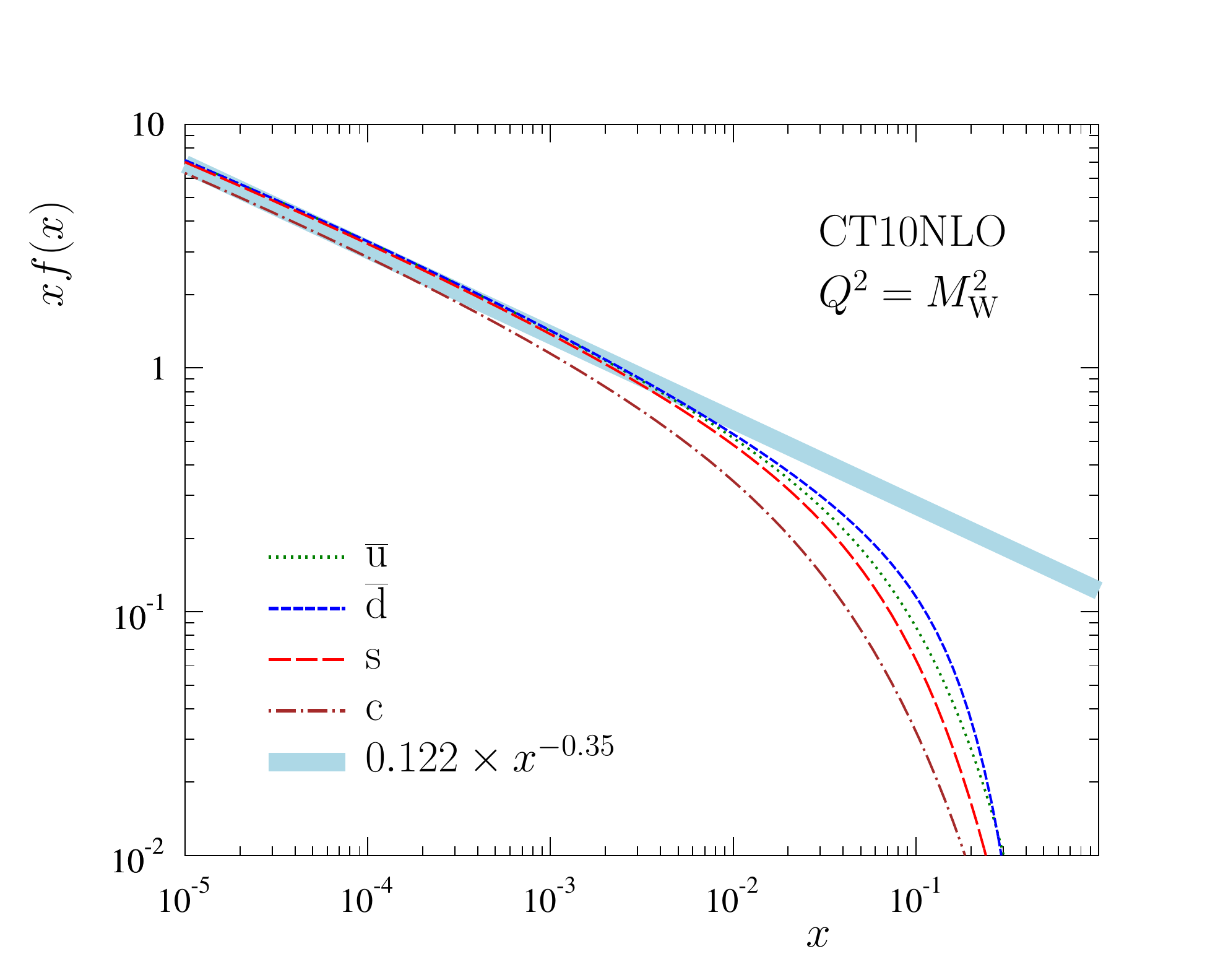}
\caption{Scaling exponent extracted from NLO calculations (upper panel) and its comparison with CT10NLO PDFs (lower panel).}
\label{fig:scalingexponent}
\end{figure} 

\begin{figure}[htb!]
\centering
\includegraphics[width=\linewidth]{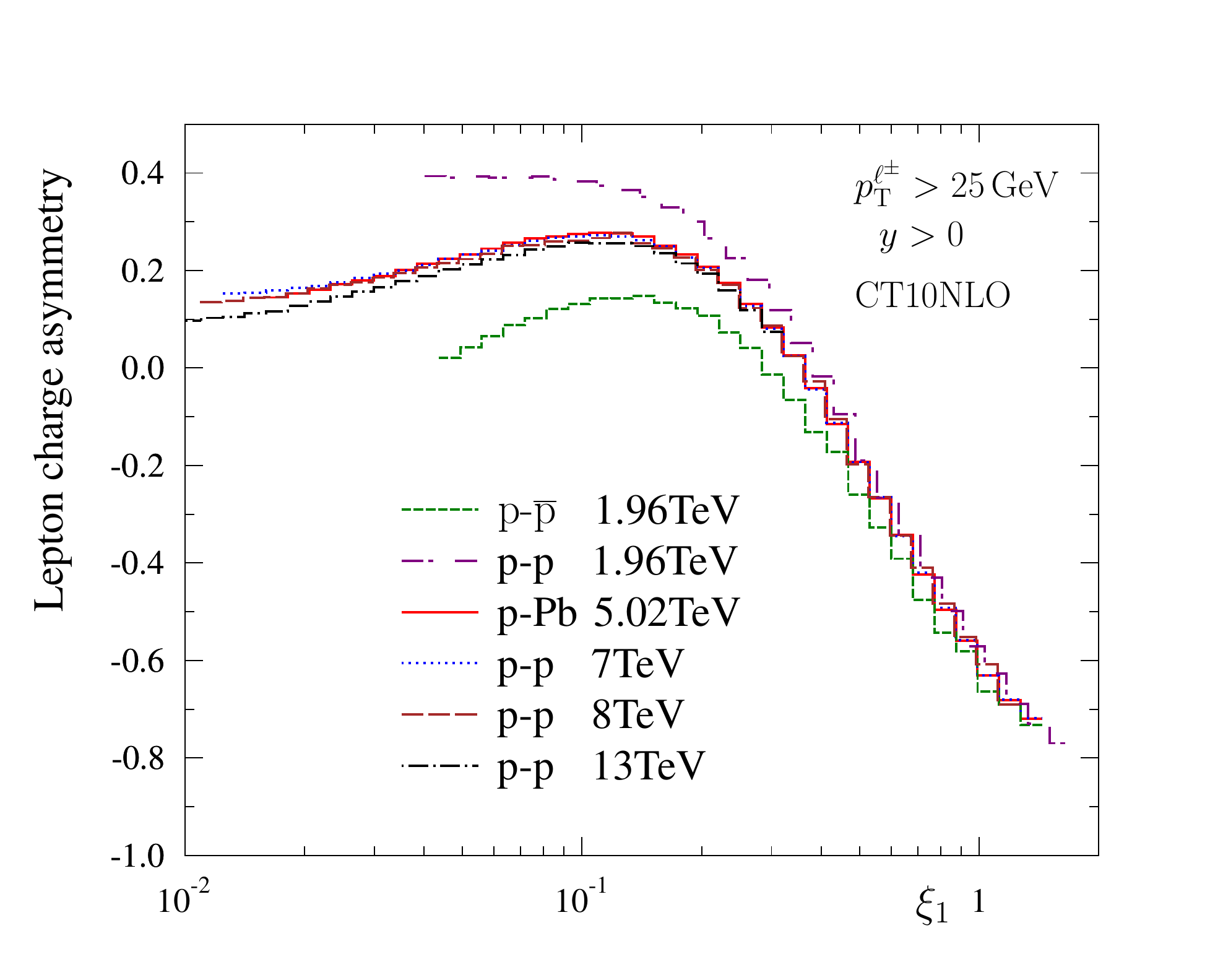}
\includegraphics[width=\linewidth]{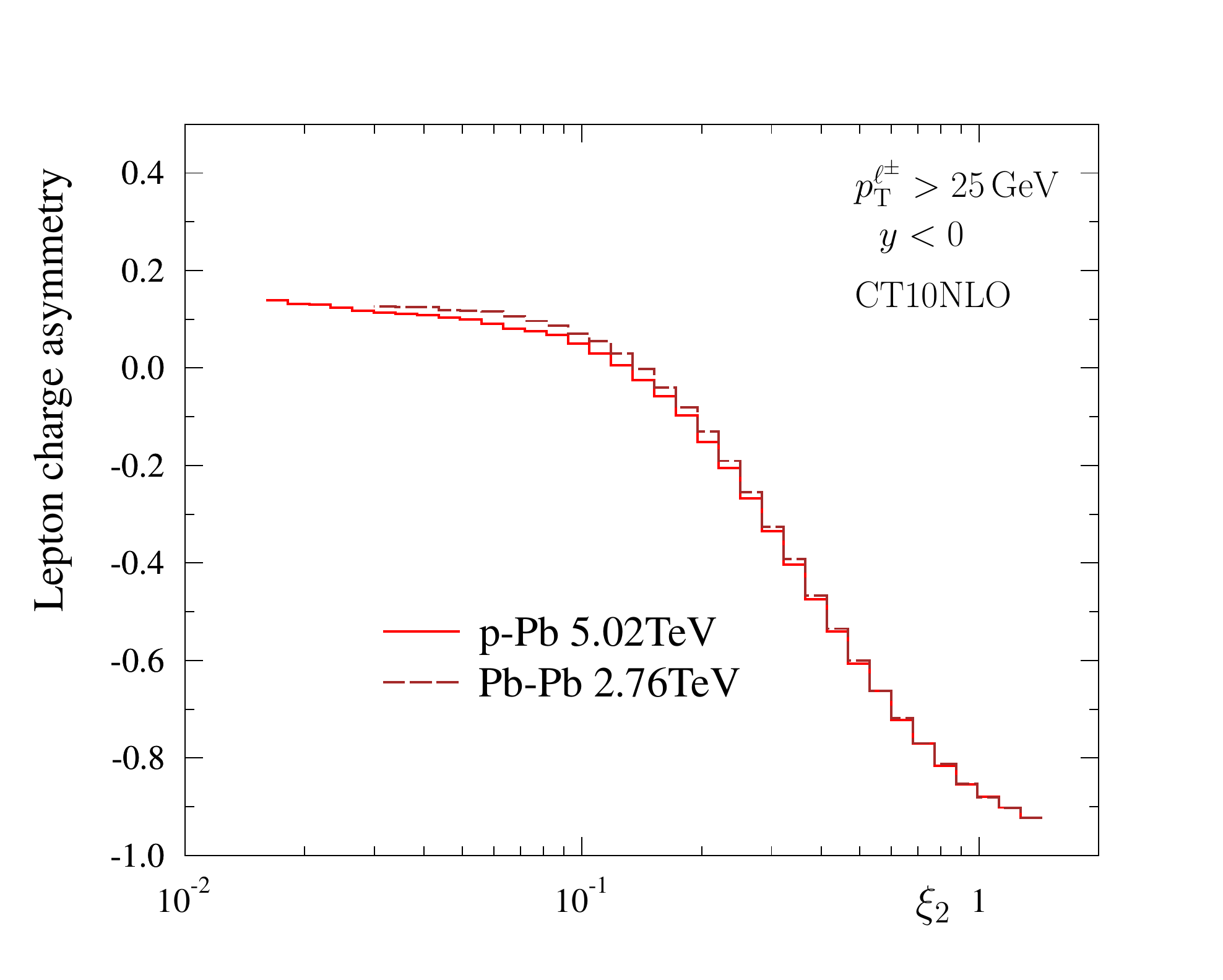}
\caption{Lepton charge asymmetry in \ppbar \,($\sqrt{s} = 1.96 \, {\rm TeV}$), \pp \,($\sqrt{s} = 1.96, 7, 8 \, {\rm TeV}$), \pPb\ \,($\sqrt{s} = 5.02 \, {\rm TeV}$) and \PbPb\ \,($\sqrt{s} = 2.76 \, {\rm TeV}$) collisions, for $y>0$ (upper panel) and $y<0$ (lower panel).}
\label{fig:theoryscaling}
\end{figure} 

According to Eq.~(\ref{eq:power}), the scaling exponent $\alpha$ in Eq.~(\ref{eq:scalingxsec}) should reflect the small-$x$ behaviour of quark distributions and it can be straightforwardly extracted from cross-sections at two different centre-of-mass energies. To verify this correspondence and the consistency of our derivation, we have computed the full NLO cross-sections at $\sqrt{s}=7,8,13 \, {\rm TeV}$ for \pp\ collisions using \texttt{MCFM} Monte-Carlo code~\cite{Campbell:2010ff} and CT10NLO PDFs~\cite{Lai:2010vv}. From these cross-sections, we have evaluated the effective scaling exponent $\alpha_{\rm eff}$ by
\begin{equation}
 \alpha_{\rm eff}(\xi) =  \log \left[ \frac{\sigma^{\ell^\pm}({s},\xi)/d\xi}{\sigma^{\ell^\pm}(s^\prime,\xi)/d\xi} \right] \log^{-1} \left( \frac{s}{{s^\prime}} \right),
\end{equation}	
taking $\sqrt{s}=7 \, {\rm TeV}$ and $\sqrt{{s^\prime}}=8,13 \, {\rm TeV}$. The outcome is plotted in the upper panel of Fig.~\ref{fig:scalingexponent}. To first approximation, towards large $\xi_{1,2}$ the effective scaling exponent is $\alpha_{\rm eff} \approx 0.35$ and independent of the lepton charge. In more detail, the scaling exponent is not exactly constant but some variation is visible which reflects the fact that the PDFs do not follow a pure power law, especially when $x$ is not very small (at small $\xi_{1,2}$). The scaling exponent for $\ell^-$ tends to have more slope and to be somewhat larger than that of $\ell^+$ especially at small $\xi_{1,2}$ which corresponds to midrapidity. This can be explained by the slightly steeper slope of the $\overline{u}$ distribution in comparison to $\overline{d}$ distribution (see the lower panel of Fig.~\ref{fig:scalingexponent}) and the fact that $\ell^-$ production tends to be sensitive to somewhat larger values of $x$ in the small-$x$ side. The latter follows from the factors $(1\pm\sqrt{1-4p_{\rm T}^2/M_{\rm W}^2})^2$ that multiply PDFs in Eq.~(\ref{eq1}). These, in turn, originate from the parity non-conserving W couplings to quarks and leptons. The lower panel in Fig.~\ref{fig:scalingexponent} compares the extracted exponent $\alpha_{\rm eff} \approx 0.35$ to the CT10NLO sea-quark PDFs. Evidently, there is a good correspondence between the scaling exponent $\alpha$ and the behaviour of the small-$x$ quark PDFs. We can conclude that despite the complex higher-order QCD calculations, the centre-of-mass dependence of the cross sections being discussed can be essentially captured by a simple power law.

Let us now discuss Eq.~(\ref{eq:masterscaling}) and whether the nature of the hadronic projectile or nucleus probed at small $x$ really disappears as conjectured. To this end we have computed the lepton charge asymmetry (again, at NLO accuracy) in various collision systems at centre-of-mass energies that correspond to existing Tevatron and  LHC data. The results are shown in Fig.~\ref{fig:theoryscaling}. At $y \gg 0$, the curves corresponding to p-p, p-Pb and p-$\overline{\rm p}$ tend to unite, whereas in the opposite direction, $y \ll 0$, p-Pb and Pb-Pb become approximately the same. Thus, as far as theoretical NLO expectations are concerned, the scaling law of Eq.~(\ref{eq:masterscaling}) turns out to be a very good approximation, though not perfect. The largest deviations in Fig.~\ref{fig:theoryscaling} are seen in the case of p-$\overline{\rm p}$ at the Tevatron energy, $\sqrt{s} = 1.96 \, {\rm TeV}$. There, the probed values of $x$ for $\overline{\rm p}$ are not small enough and especially the assumption of charge-conjugation symmetric quark distributions, Eq.~(\ref{eq:seaquark}), is not particularly accurate until almost the end of phase space
(e.g. $\xi_1=1$ corresponds to $x_2 \approx M_{\rm W}^2/s \approx 0.002$). The p-p curve at the same center-of-mass energy unites with the rest already at lower $\xi_1$.

At small fixed value of $\xi$, the lepton charge asymmetry in p-p collisions tends to decrease towards increasing centre-of-mass energies. This can be interpreted in terms of slightly different scaling exponent for $\ell^+$ and $\ell^-$ production (see Fig.~\ref{fig:scalingexponent}). Denoting the scaling exponent for $\ell^\pm$ production by $\alpha^\pm$, and the difference by $\Delta \equiv \alpha^- - \alpha^+$, to first approximation, \vspace{0.2cm}
\begin{eqnarray}
& & {\mathcal C}_\ell^{{\rm H}_1,{\rm H}_2}(s',\xi) =  {\mathcal C}_\ell^{{\rm H}_1,{\rm H}_2}(s,\xi) \label{eq:differents} \\
& & + \frac{\Delta}{2}
\left\{1 - \left[{\mathcal C}_\ell^{{\rm H}_1,{\rm H}_2}(s,\xi)\right]^2 \right\}
\log\left(\frac{s}{s'}\right)  + \mathcal{O} \left( \Delta^2\right). \nonumber
\end{eqnarray}
Since $\Delta > 0$, we have a condition
\begin{equation}
{\mathcal C}_\ell^{{\rm H}_1,{\rm H}_2}(s',\xi) < {\mathcal C}_\ell^{{\rm H}_1,{\rm H}_2}(s,\xi), \,\, {\rm if} \,\, s' > s,
\end{equation}
which explains the decreasing trend of lepton charge asymmetries in p-p collisions towards higher centre-of-mass energies at fixed, small $\xi$.

\section{Data and predictions}
\label{Dataandpredictions}

\subsection{Comparison with existing data}

\begin{table*}
\center
\caption{The experimental data sets.}
\label{tab:Data_and_cuts}
\begin{tabular*}{\textwidth}{@{\extracolsep{\fill}}ccccc@{}}
  Experiment & System & $\sqrt{s}$ & kinematic cuts & Ref.\\
  \hline
  D0      & p-$\overline{\rm p}$    & 1.96 TeV & $p_{\rm T} > 25 \, {\rm GeV}$, $\slashed{E}_{\rm T} > 25 \, {\rm GeV}$ &~\cite{D0:2014kma} \\
  ATLAS      & Pb-Pb    & 2.76 TeV & $p_{\rm T} > 25 \, {\rm GeV}$, $\slashed{E}_{\rm T} > 25 \, {\rm GeV}$, $m_{\rm T} > 40 \, {\rm GeV}$ &~\cite{Aad:2014bha} \\
  CMS        & p-Pb    & 5.02 TeV & $p_{\rm T} > 25 \, {\rm GeV}$ &~\cite{Khachatryan:2015hha} \\
  ALICE      & p-Pb    & 5.02 TeV & $p_{\rm T} > 10 \, {\rm GeV}$ &~\cite{Zhu:2015kpa} \\
  CMS        & \pp\     & 7 TeV & $p_{\rm T} > 25 \, {\rm GeV}$ &~\cite{Chatrchyan:2013mza} \\
  ATLAS      & \pp\     & 7 TeV & $p_{\rm T} > 20 \, {\rm GeV}$, $\slashed{E}_{\rm T} > 25 \, {\rm GeV}$, $m_{\rm T} > 40 \, {\rm GeV}$ &~\cite{Aad:2011dm} \\
  LHCb       & \pp\     & 7 TeV & $p_{\rm T} > 20 \, {\rm GeV}$ &~\cite{Aaij:2014wba} \\
  LHCb       & \pp\     & 8 TeV & $p_{\rm T} > 20 \, {\rm GeV}$ &~\cite{Aaij:2015zlq} \\
  CMS        & \pp\     & 8 TeV & $p_{\rm T} > 25 \, {\rm GeV}$ &~\cite{Khachatryan:2016pev} \\
 \end{tabular*}
\end{table*}

\begin{figure*}[htb!]
\centering
\includegraphics[width=0.48\linewidth]{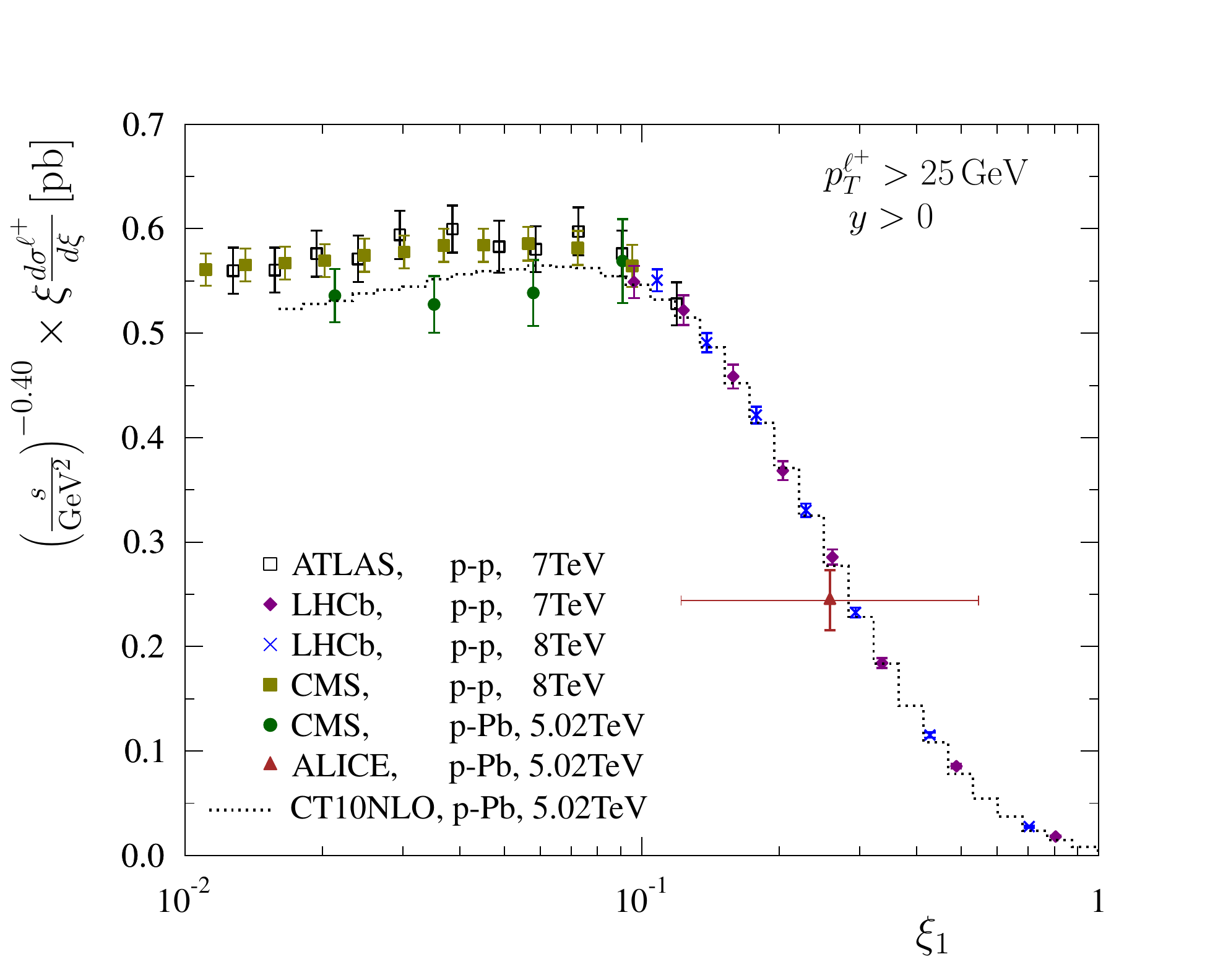}
\includegraphics[width=0.48\linewidth]{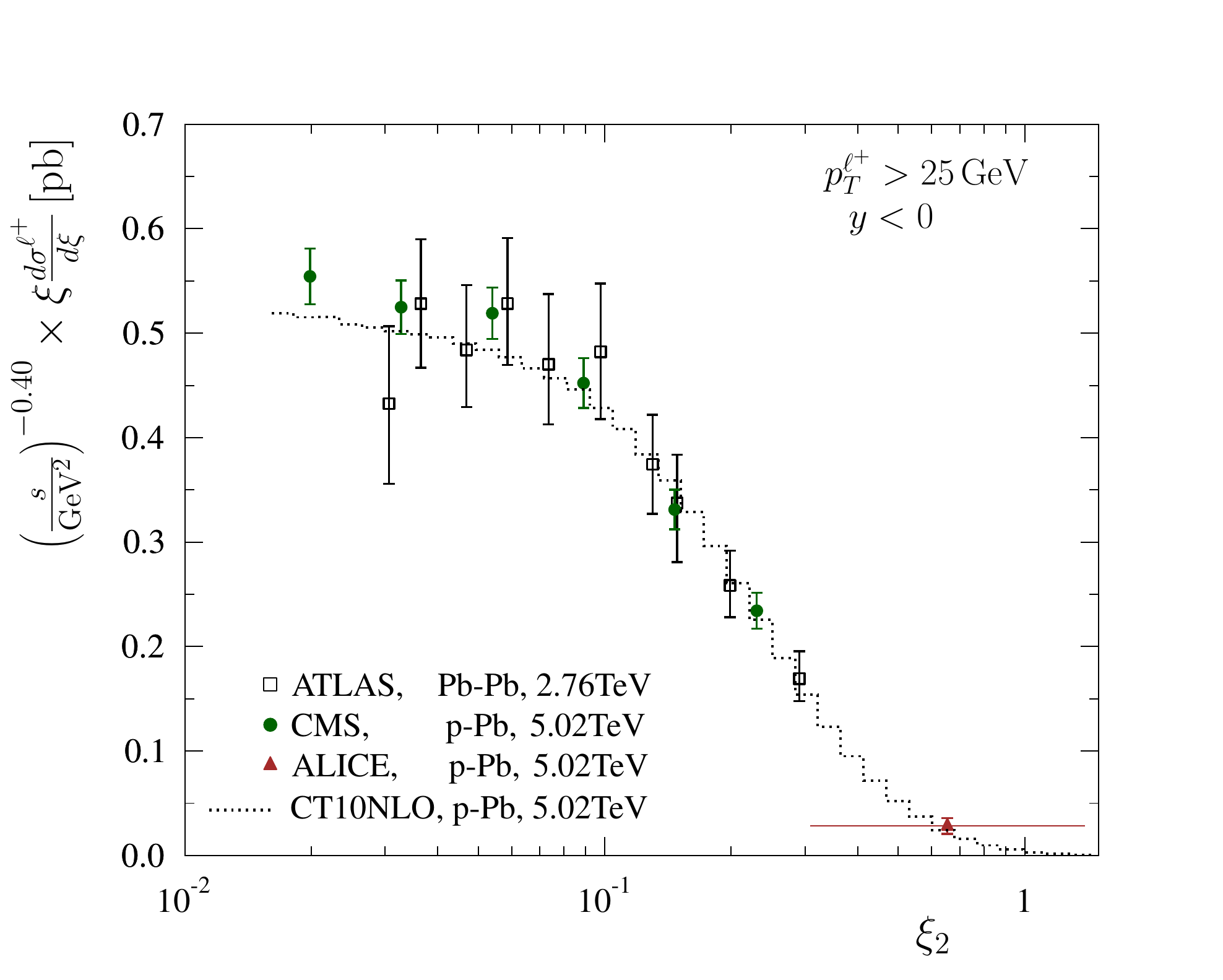}
\includegraphics[width=0.48\linewidth]{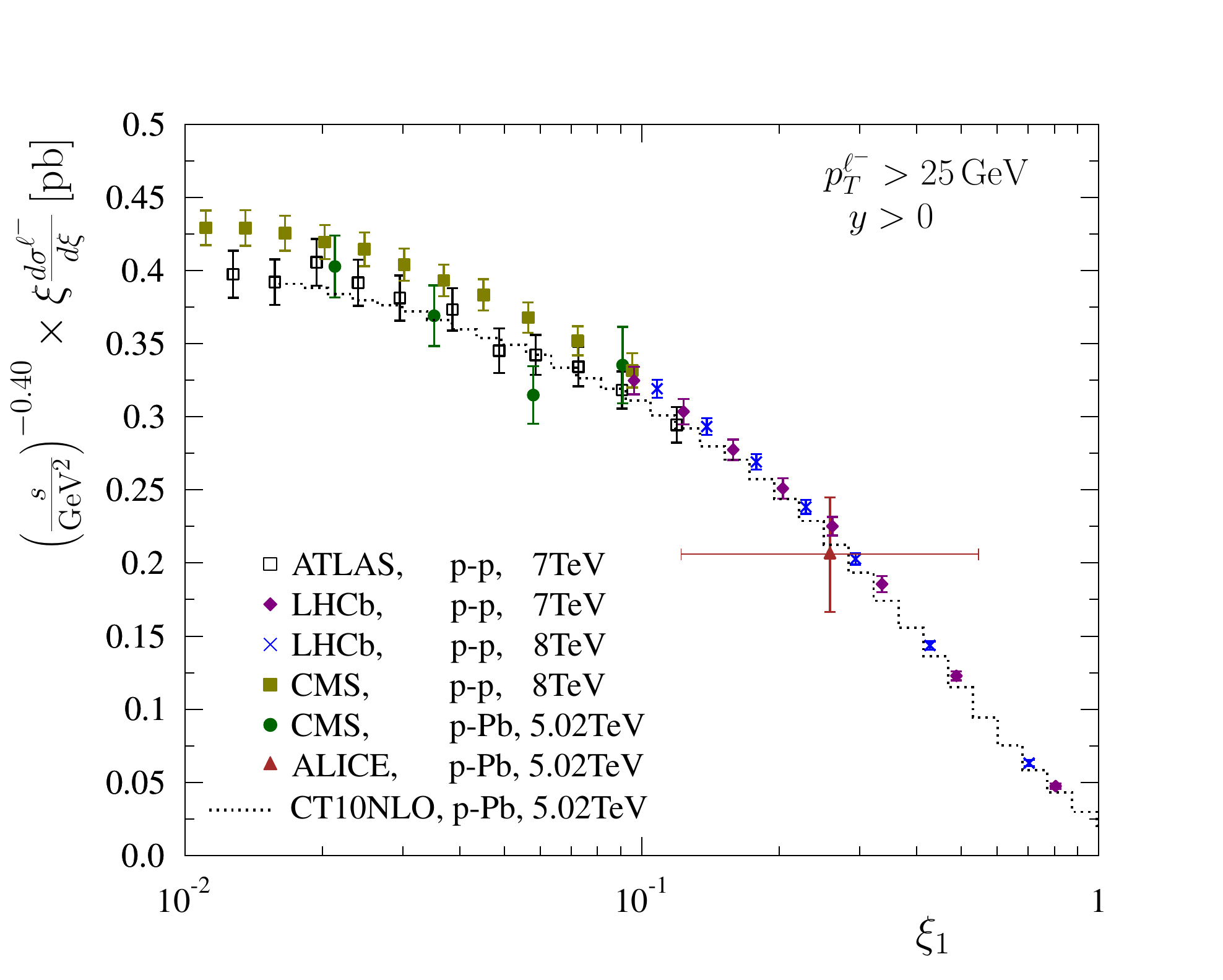}
\includegraphics[width=0.48\linewidth]{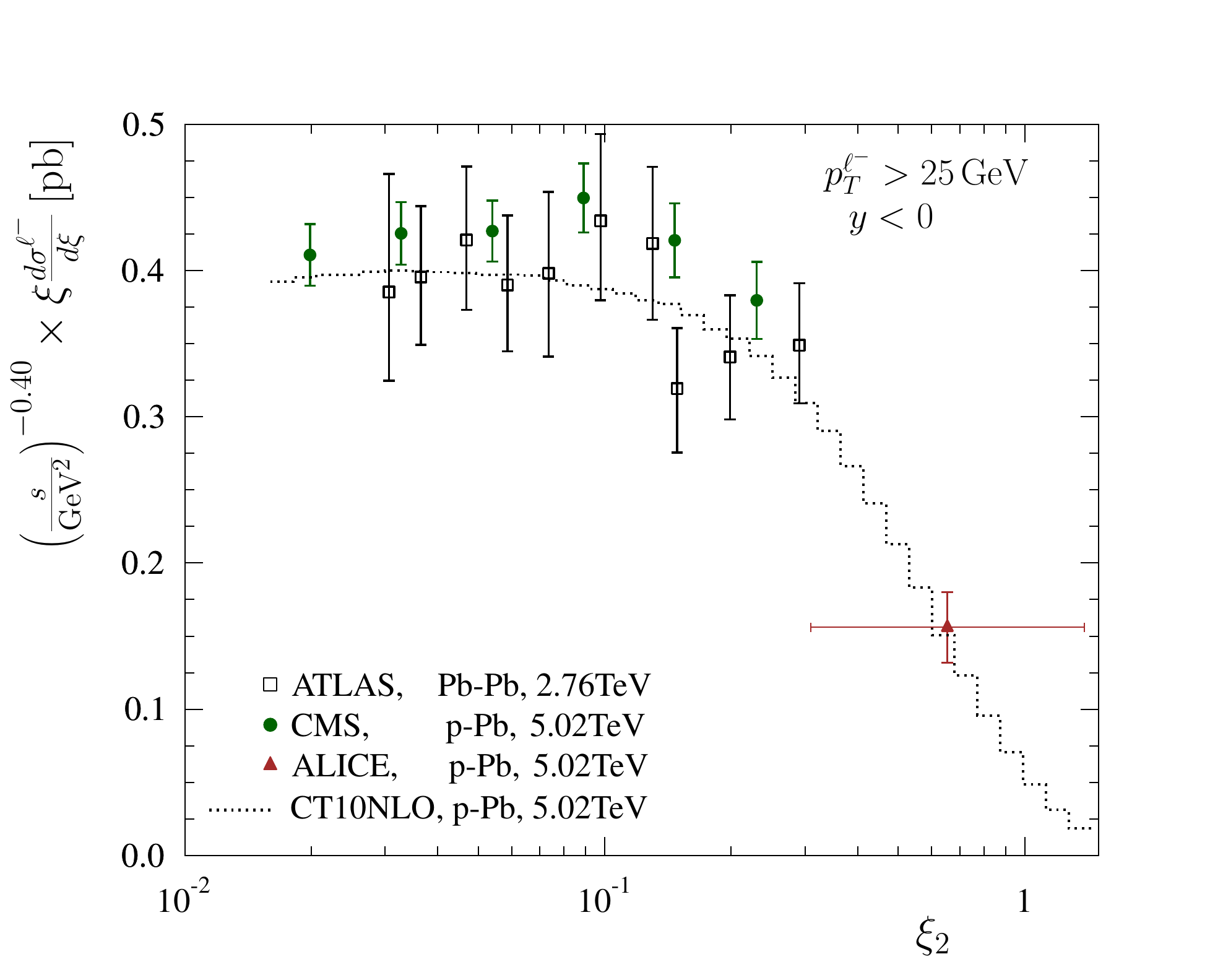}
\caption{Absolute spectra of charged leptons (upper panels for $\ell^+$, lower panels for $\ell^-$) in p-p \,($\sqrt{s} = 7, 8 \, {\rm TeV}$) and \pPb\ \,($\sqrt{s} = 5.02 \, {\rm TeV}$) collisions for $y>0$ (left-hand panles), and in Pb-Pb \,($\sqrt{s} = 2.76 \, {\rm TeV}$) and \pPb\ \,($\sqrt{s} = 5.02 \, {\rm TeV}$) collisions for $y<0$. The data has been scaled by $(s/{\rm GeV}^2)^{-0.40}$.}
\label{fig:absolutescaling}
\end{figure*} 


\begin{figure*}[htb!]
\centering
\includegraphics[width=0.65\linewidth]{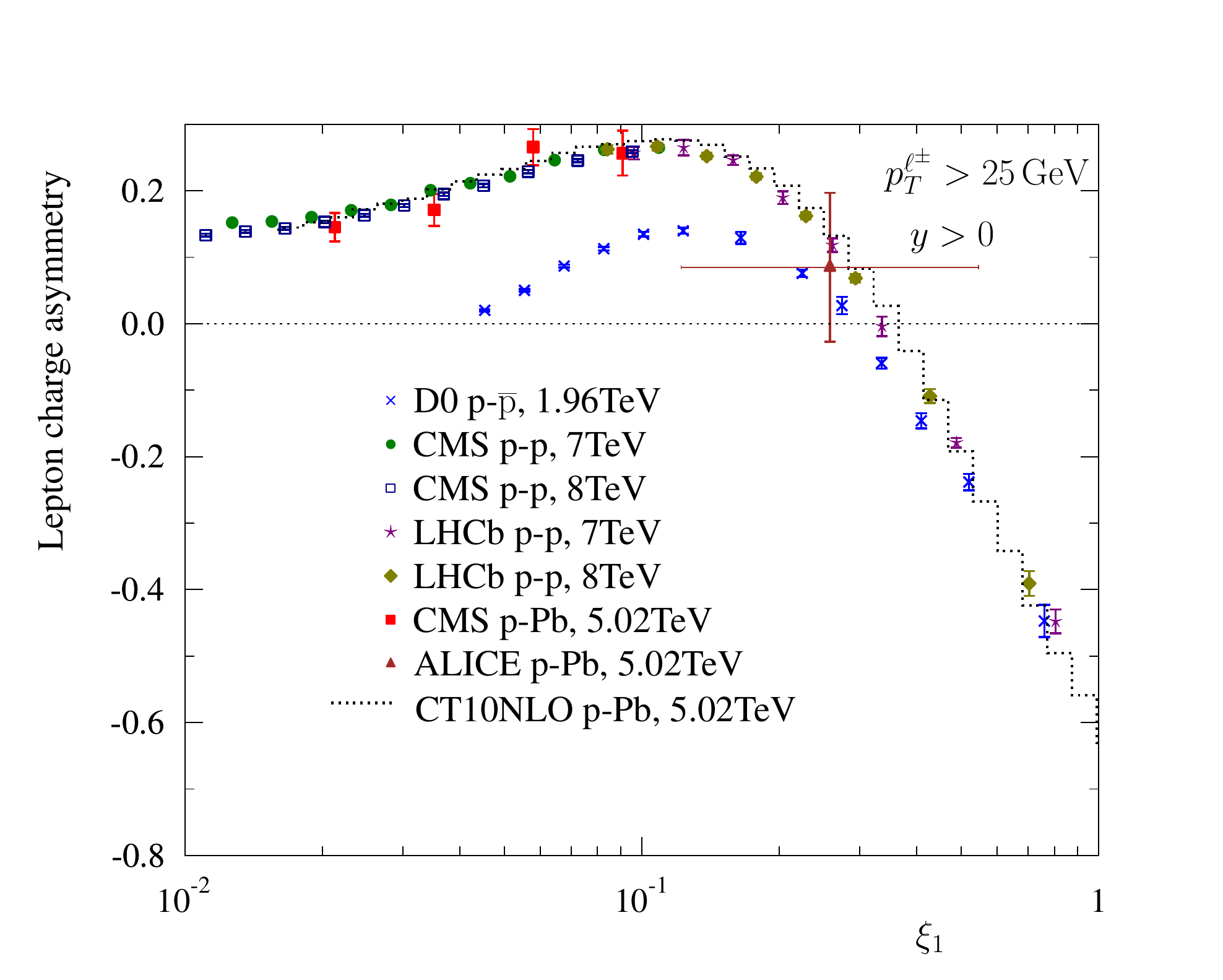}\vspace{-1cm}
\includegraphics[width=0.65\linewidth]{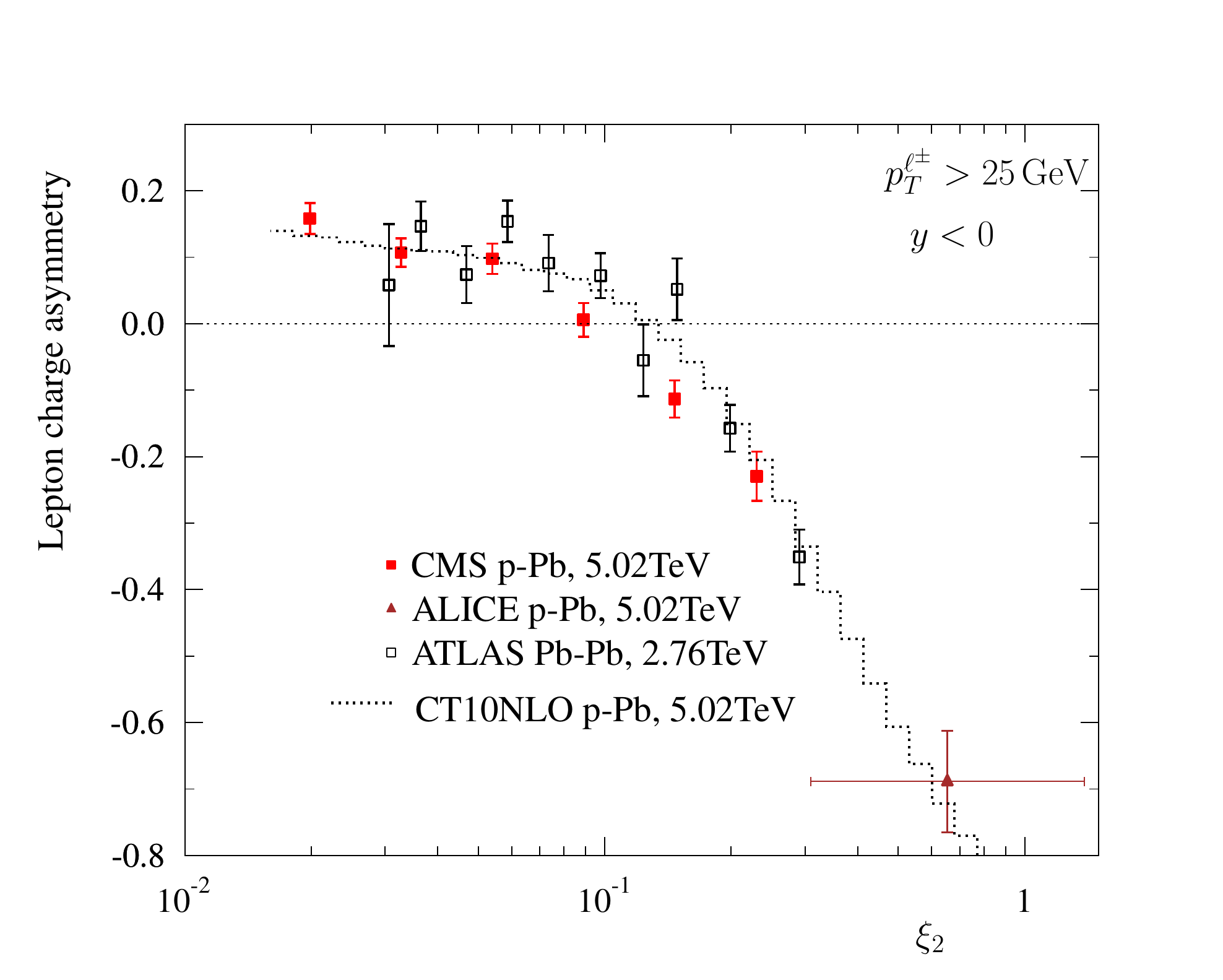}
\caption{Lepton charge asymmetry in \ppbar \,($\sqrt{s} = 1.96 \, {\rm TeV}$), \pp \,($\sqrt{s} = 7,8 \, {\rm TeV}$), \pPb\ \,($\sqrt{s} = 5.02 \, {\rm TeV}$) and \PbPb\ \,($\sqrt{s} = 2.76 \, {\rm TeV}$) collisions. The dotted curve is to guide the eye and corresponds to $\mathcal{C}_\ell^{\rm p,Pb}$ at $\sqrt{s} = 5.02 \, {\rm TeV}$.}
\label{fig:theoryscalingHP}
\end{figure*} 

\begin{figure*}[htb!]
\centering
\includegraphics[width=0.8\linewidth]{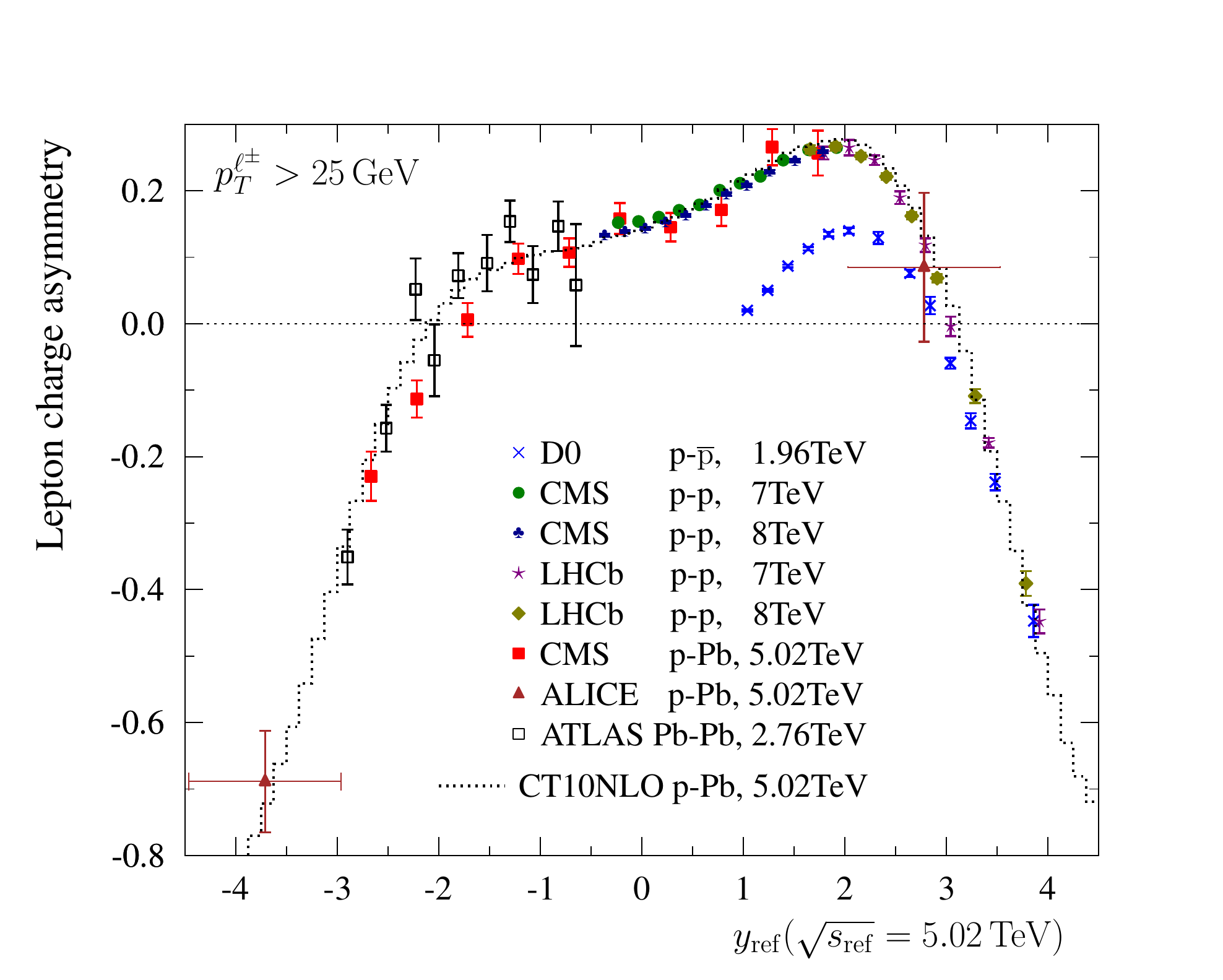}
\caption{The world data on lepton charge asymmetry as a function of $y_{\rm ref}$ taking $\sqrt{s_{\rm ref}} = 5.02 \, {\rm TeV}$.}
\label{fig:masterplot}
\end{figure*} 

The currently most accurate experimental measurements for inclusive W production from Tevatron and LHC experiments are summarized in Table~\ref{tab:Data_and_cuts}. A direct comparison of various measurements is complicated by the kinematic cuts for lepton $p_{\rm T}$, missing transverse energy $\slashed{E}_{\rm T}$, and transverse mass $m_{\rm T}$ of the neutrino-lepton system, which vary among the experiments and have to be accounted for. Here, we have chosen to ``correct'' the data to $p_{\rm T} > 25 \, {\rm GeV}$ (the default cut in CMS measurements) by \texttt{MCFM} evaluating the observables first with the true cuts shown in Table~\ref{tab:Data_and_cuts}, then with $p_{\rm T} > 25 \, {\rm GeV}$ and taking the ratio (absolute cross sections) or difference (charge asymmetry). We stress that if the kinematic cuts were the same in all experiments, this step would be unnecessary. The available absolute cross sections are compared in Fig.~\ref{fig:absolutescaling}. The p-p and p-Pb data are plotted together at forward rapidity (left-hand panels) and Pb-Pb and p-Pb data together at backward rapidity (right-hand panels). In these plots, the data has been scaled by a factor $(s/{\rm GeV}^2)^{-\alpha}$, 
where a constant value $\alpha=0.4$ has been used for the scaling exponent as a compromise between the expected exponent at small and large $\xi$, see Fig.~\ref{fig:scalingexponent}. Keeping in mind the ``non-constantness'' of the scaling exponent and that at forward (backward) direction the p-Pb (Pb-Pb) data are presumably affected by small-$x$ shadowing in comparison to p-p (p-Pb), an exact match with p-p (p-Pb) is not expected. Nevertheless, there is clearly a rough correspondence between the data from different collision systems and different $\sqrt{s}$.

The data for lepton charge asymmetries $\mathcal{C}_\ell$ are compiled in Fig.~\ref{fig:theoryscalingHP}. We note that some experimental uncertainties, luminosity above all, cancel in the measurement of the lepton charge asymmetries as compared to absolute cross sections. As previously, the data from p-p, p-$\overline{\rm p}$, and Pb-Pb collisions are plotted only in the direction where they are supposed to merge with p-Pb data. To a very good approximation, the experimental data indeed line up to the same underlying curve which corresponds to the charge asymmetry in p-Pb collisions. Two CMS p-Pb data points at negative rapidities appear to lie below the NLO predictions and could potentially require additional nuclear modifications in PDFs (as also pointed out in Ref.~\cite{Khachatryan:2015hha}). However, the ATLAS Pb-Pb data shows no sign of such a disagreement with the theory at those values of rapidity indicating that there appears to be some tension between these two data sets and that the both data sets cannot be optimally reproduced with the same set of (nuclear) PDFs.

We can also compress all the data into a single plot. This is done by choosing a certain reference centre-of-mass energy $\sqrt{s_{\rm ref}}$ (we take $\sqrt{s_{\rm ref}} = 5.02 \, {\rm TeV}$) and plotting the data as a function of variable
\begin{equation}
y_{\rm ref} \equiv y \pm  \frac{1}{2}\log \frac{s_{\rm ref}}{s}, \quad y \gtrless 0,
\end{equation}
such that 
\begin{eqnarray}
\xi_1(y,\sqrt{s}) & = & \xi_1(y_{\rm ref},\sqrt{s_{\rm ref}}), \quad y > 0, \\
\xi_2(y,\sqrt{s}) & = & \xi_2(y_{\rm ref},\sqrt{s_{\rm ref}}), \quad y < 0. \nonumber
\end{eqnarray}
Such a plot is shown in Fig.~\ref{fig:masterplot}. In order to keep the plot readable Pb-Pb data is plotted only at $y<0$, and p-p, p-$\overline{\rm p}$ data is plotted only at $y>0$.

\subsection{Cross-section ratios}
\label{Crosssectionratios}

In Section \ref{sec:AbsScalingproperties} we noted that ratios of cross-sections at two nearby $\sqrt{s}$ at fixed values of scaling variable $\xi_{1,2}$ could become less prone to large-$x$ PDF uncertainties in comparison to taking the ratios at fixed rapidity. To investigate this statement quantitatively, we have computed (p-p collisions, NLO precision) ratios
\begin{eqnarray}
R_{\sqrt{s'}/\sqrt{s}}^+ (y_{\rm ref}) & = & \frac{d\sigma^{\rm \ell^+}(\sqrt{s'})/dy_{\rm ref}}{d\sigma^{\rm \ell^+}(\sqrt{s})/dy_{\rm ref}} \approx \left( \frac{\sqrt{s'}}{\sqrt{s}} \right)^{2\alpha}, \\
R_{\sqrt{s'}/\sqrt{s}}^- (y_{\rm ref}) & = & \frac{d\sigma^{\rm \ell^-}(\sqrt{s'})/dy_{\rm ref}}{d\sigma^{\rm \ell^-}(\sqrt{s})/dy_{\rm ref}} \approx \left( \frac{\sqrt{s'}}{\sqrt{s}} \right)^{2\alpha}, \\
R_{\sqrt{s'}/\sqrt{s}}   (y_{\rm ref}) & = & \frac{R_{\sqrt{s'}/\sqrt{s}}^+(y_{\rm ref})}{R_{\sqrt{s'}/\sqrt{s}}^-(y_{\rm ref})} \approx 1, \label{eq:doubleratio}
\end{eqnarray}
where the prediction from scaling laws are also indicated. For comparison we evaluate the same ratios also at fixed rapidity (instead of fixed $y_{\rm ref}$). We have used \texttt{PDF4LHC15\_30NLO} set of PDFs \cite{Butterworth:2015oua} available from the LHAPDF libarary \cite{Buckley:2014ana}. This is a hybrid set that combines \cite{Gao:2013bia} information from independent PDF fits (CT14 \cite{Dulat:2015mca}, MMHT14 \cite{Harland-Lang:2014zoa}, NNPDF3.0 \cite{Ball:2014uwa}) thereby giving a better idea of the uncertainties than when sticking to a one particular PDF provider.

\begin{figure*}[htb!]
\centering
\includegraphics[width=0.32\linewidth]{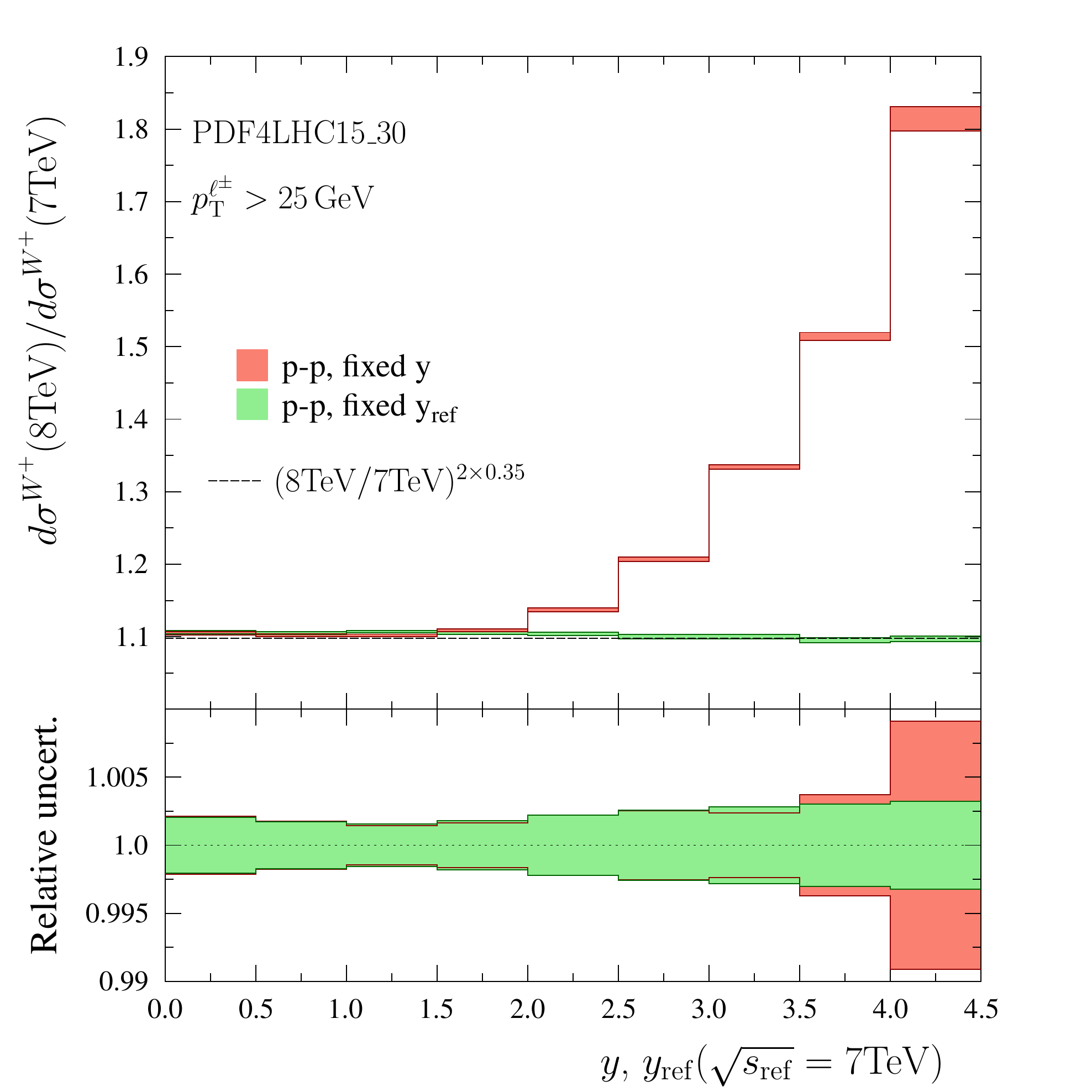}
\includegraphics[width=0.32\linewidth]{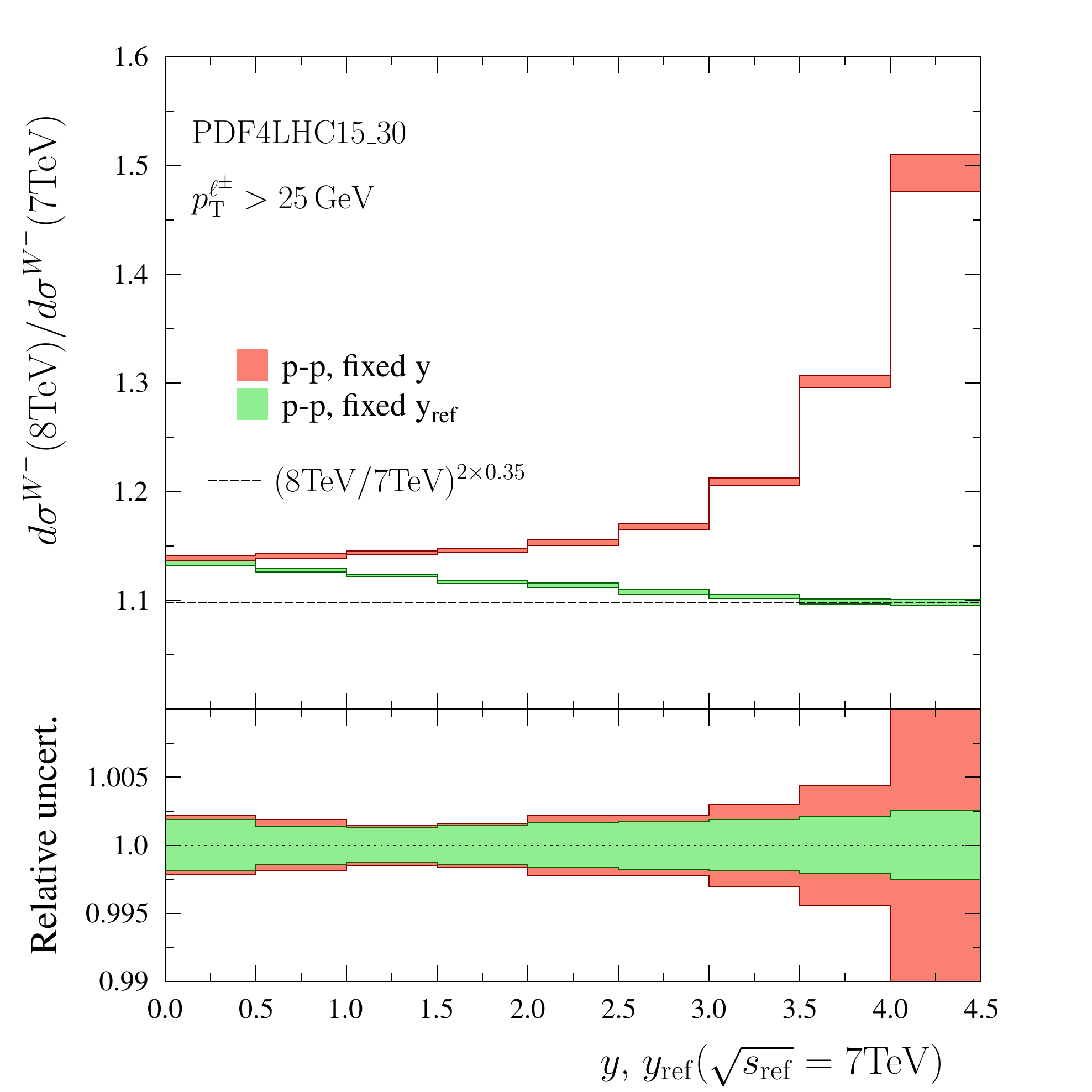}
\includegraphics[width=0.32\linewidth]{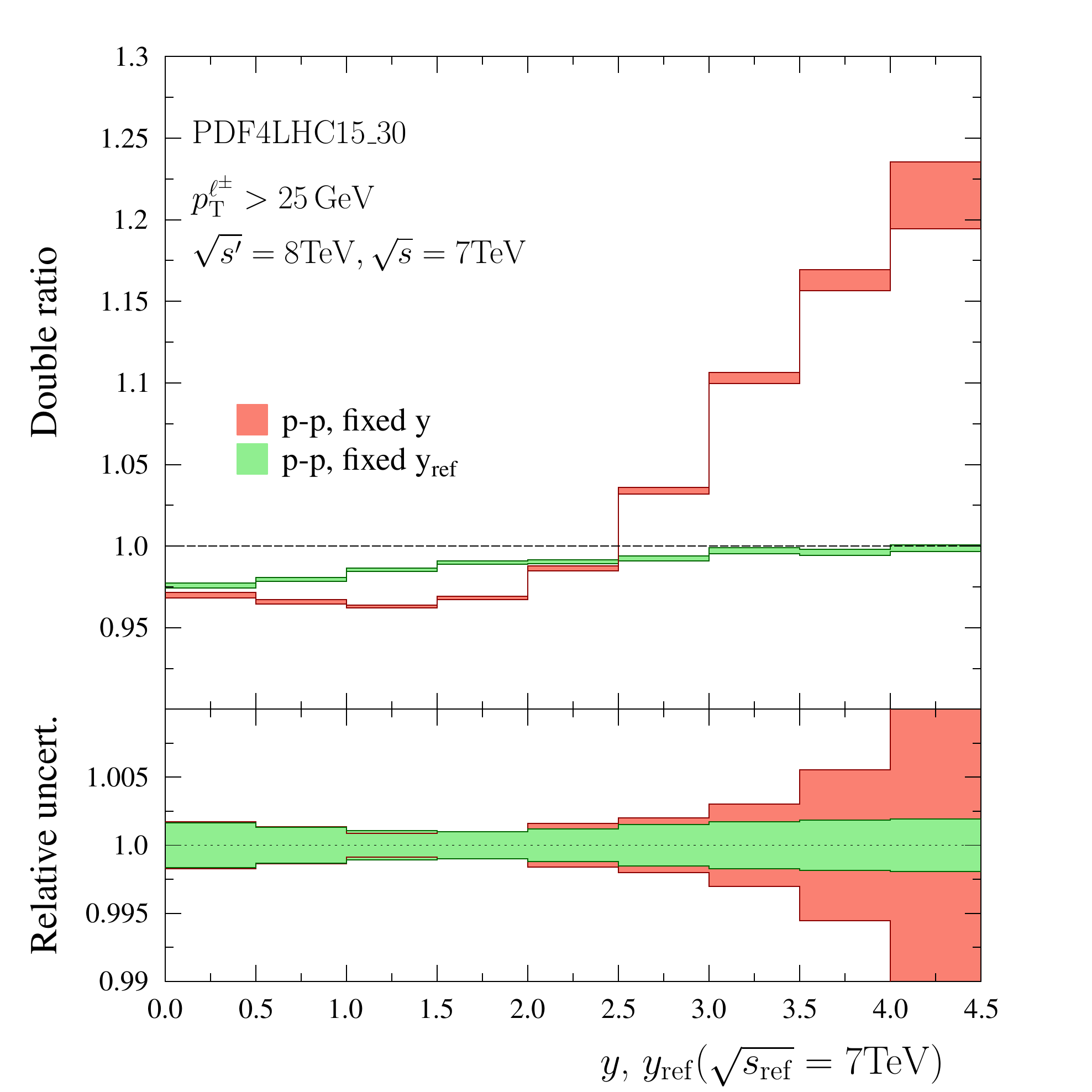}
\caption{Ratios of $\ell^+$ (left) and $\ell^-$ (middle) spectra computed at $\sqrt{s}=8\,{\rm TeV}$ and $\sqrt{s}=7\,{\rm TeV}$ center-of-mass energies. In red color are the results binned in lepton rapidity $y$, and in green the results binned in $y_{\rm ref}$ taking $\sqrt{s_{\rm ref}}=7\,{\rm TeV}$. The dashed lines indicate the prediction of scaling law Eq.~(\ref{eq:scalingxsec}). The right-hand panel shows the double ratio of Eq.~(\ref{eq:doubleratio}). }
\label{fig:ratio_8_7}
\end{figure*} 

\begin{figure*}[htb!]
\centering
\includegraphics[width=0.32\linewidth]{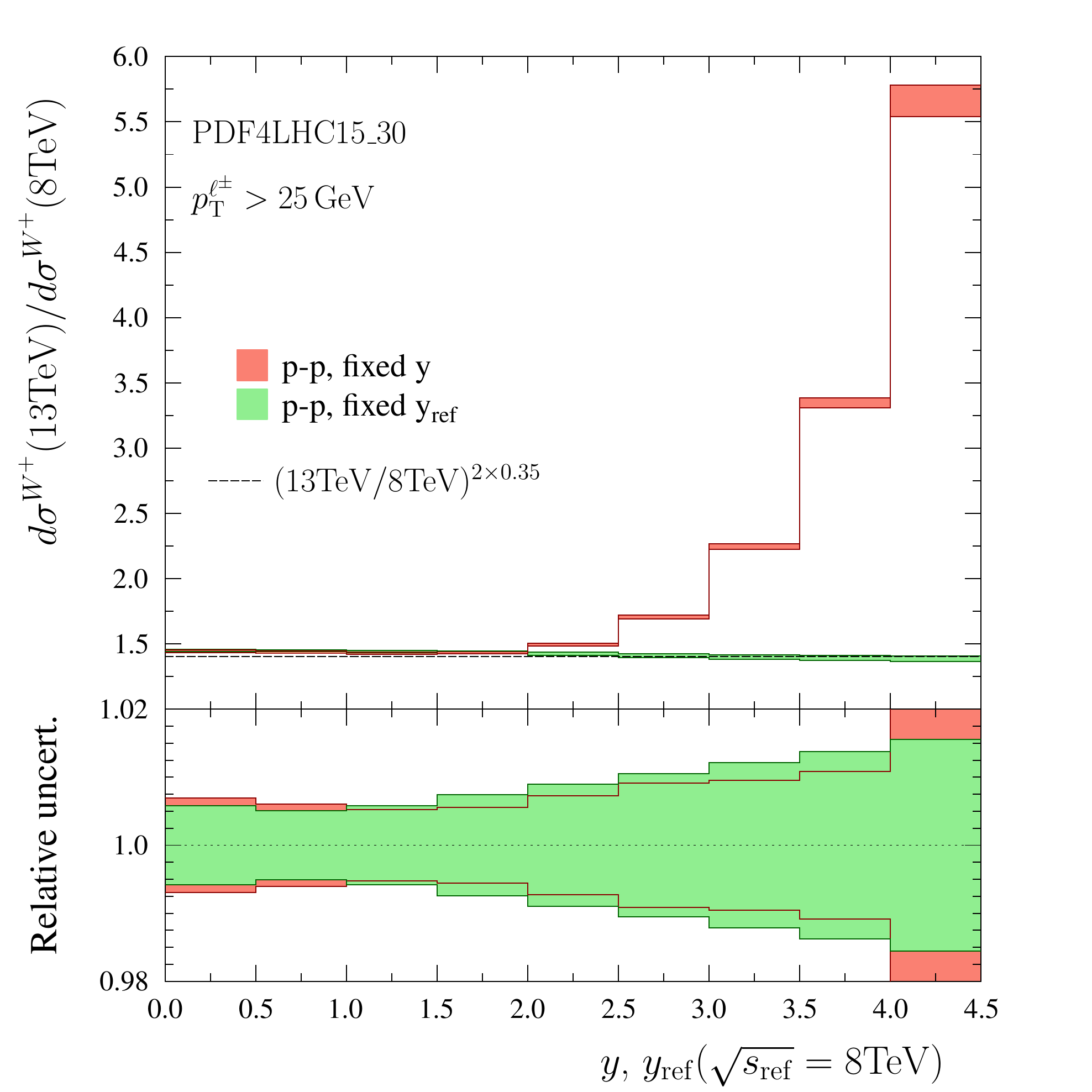}
\includegraphics[width=0.32\linewidth]{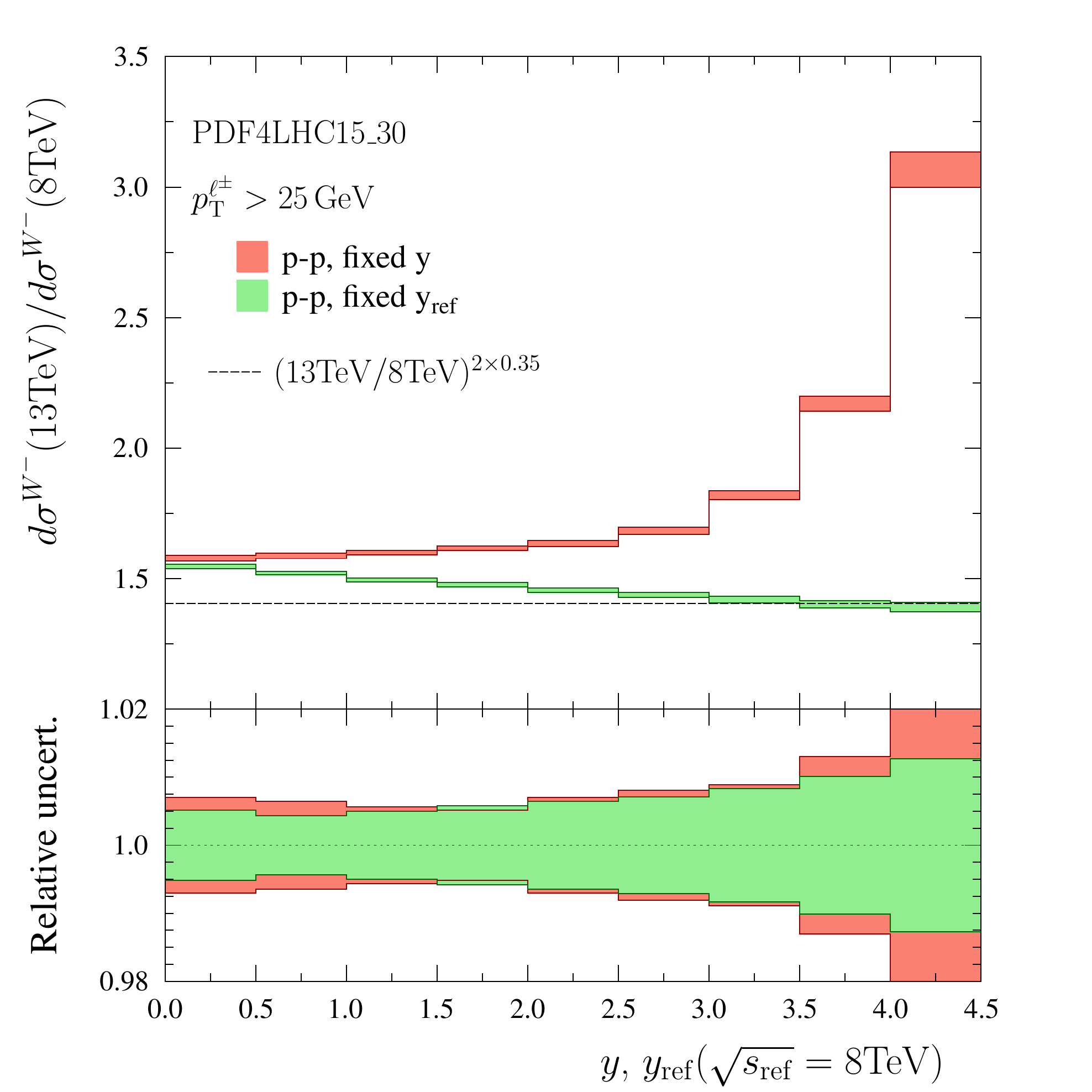}
\includegraphics[width=0.32\linewidth]{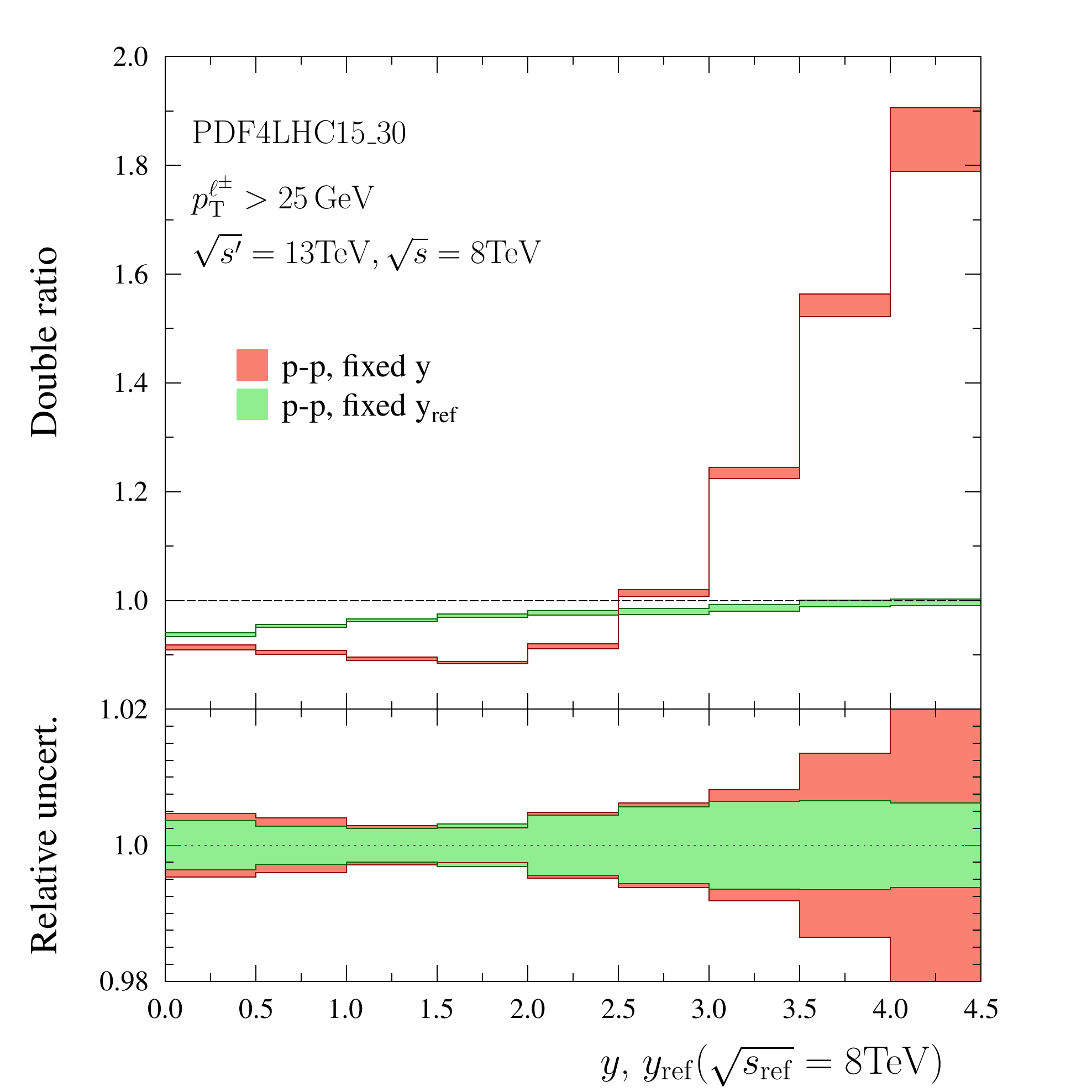}
\caption{As Fig.~\ref{fig:ratio_8_7} but using $\sqrt{s'}=13\,{\rm TeV}$ and $\sqrt{s}=8\,{\rm TeV}$.}
\label{fig:ratio_13_8}
\end{figure*} 

\begin{figure*}[htb!]
\centering
\includegraphics[width=0.32\linewidth]{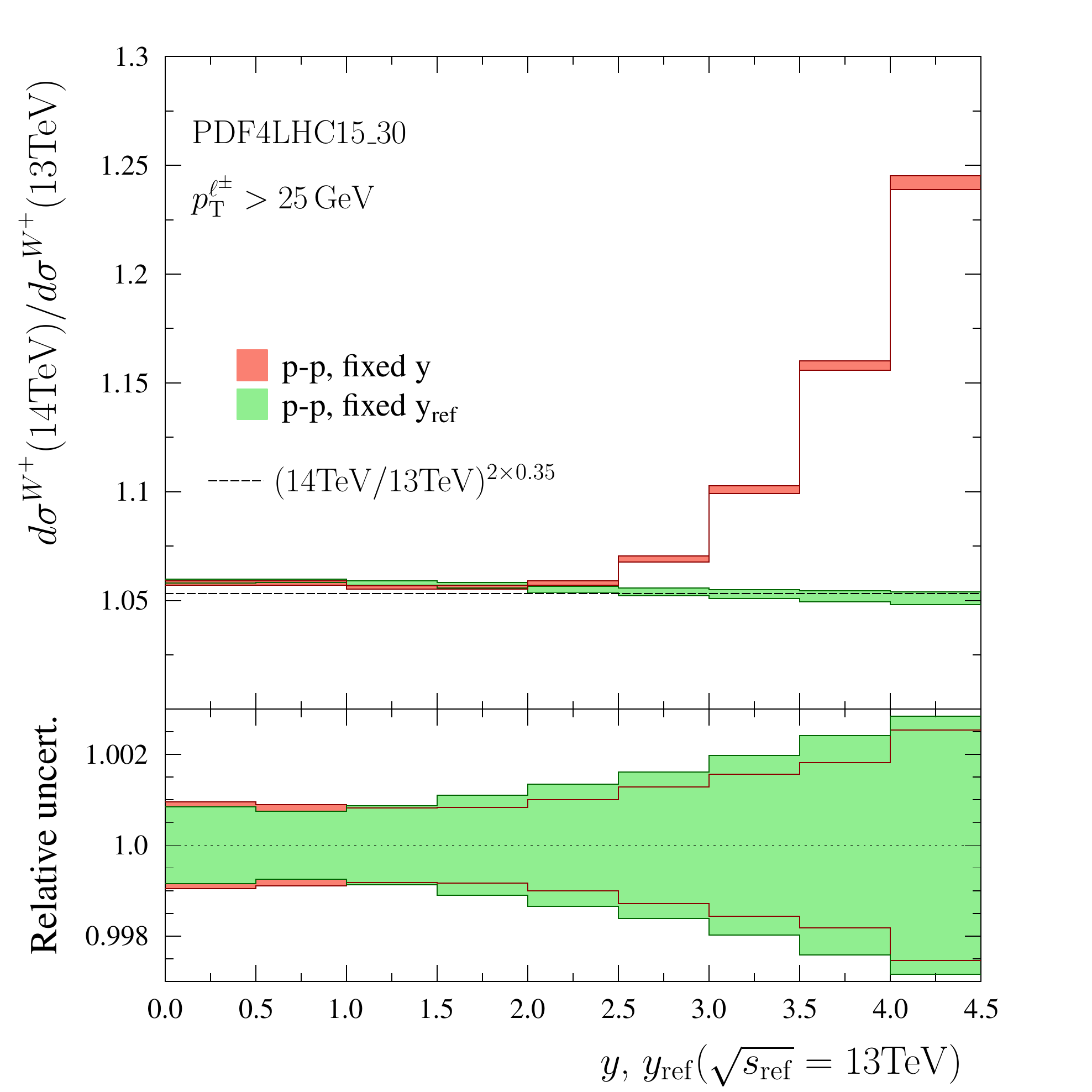}
\includegraphics[width=0.32\linewidth]{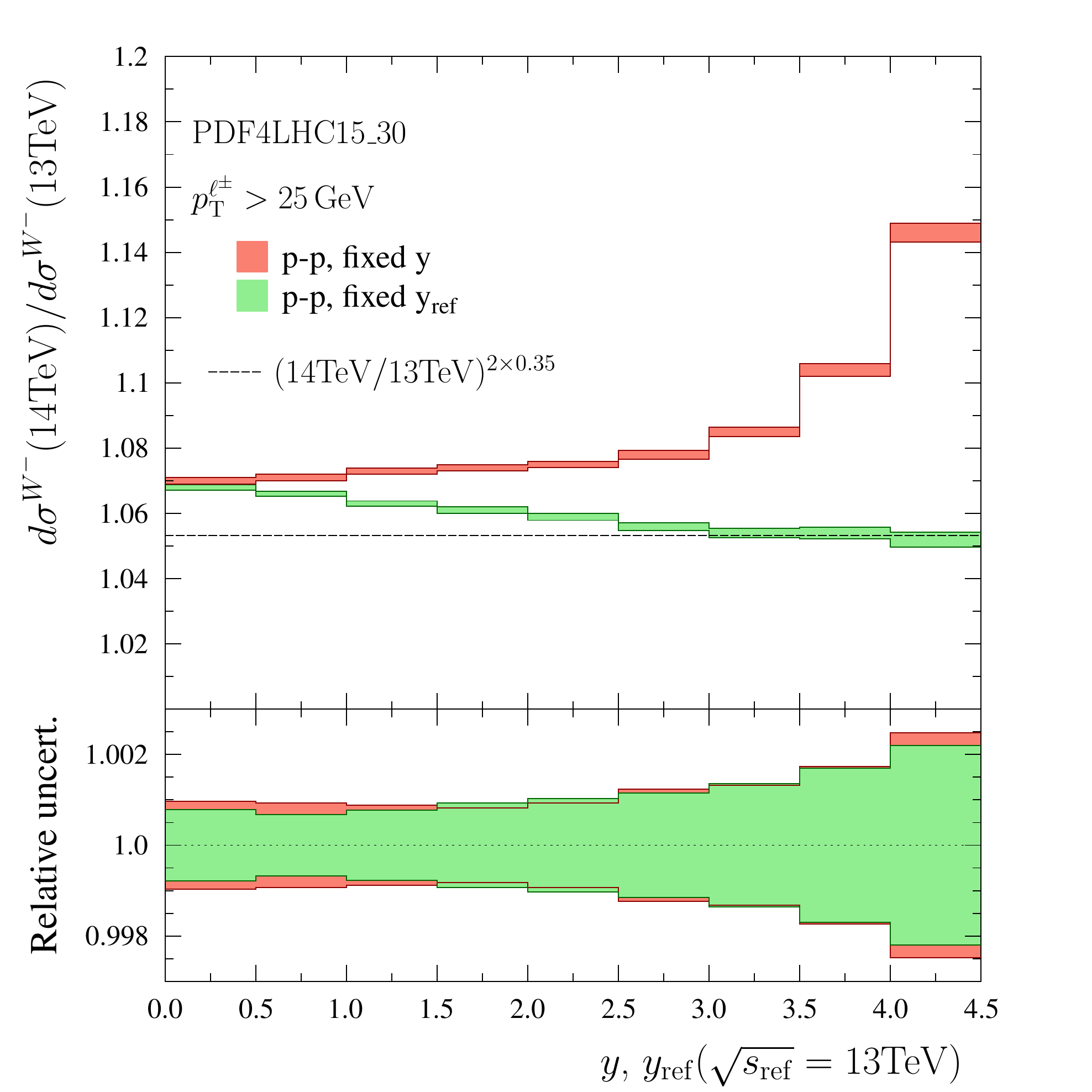}
\includegraphics[width=0.32\linewidth]{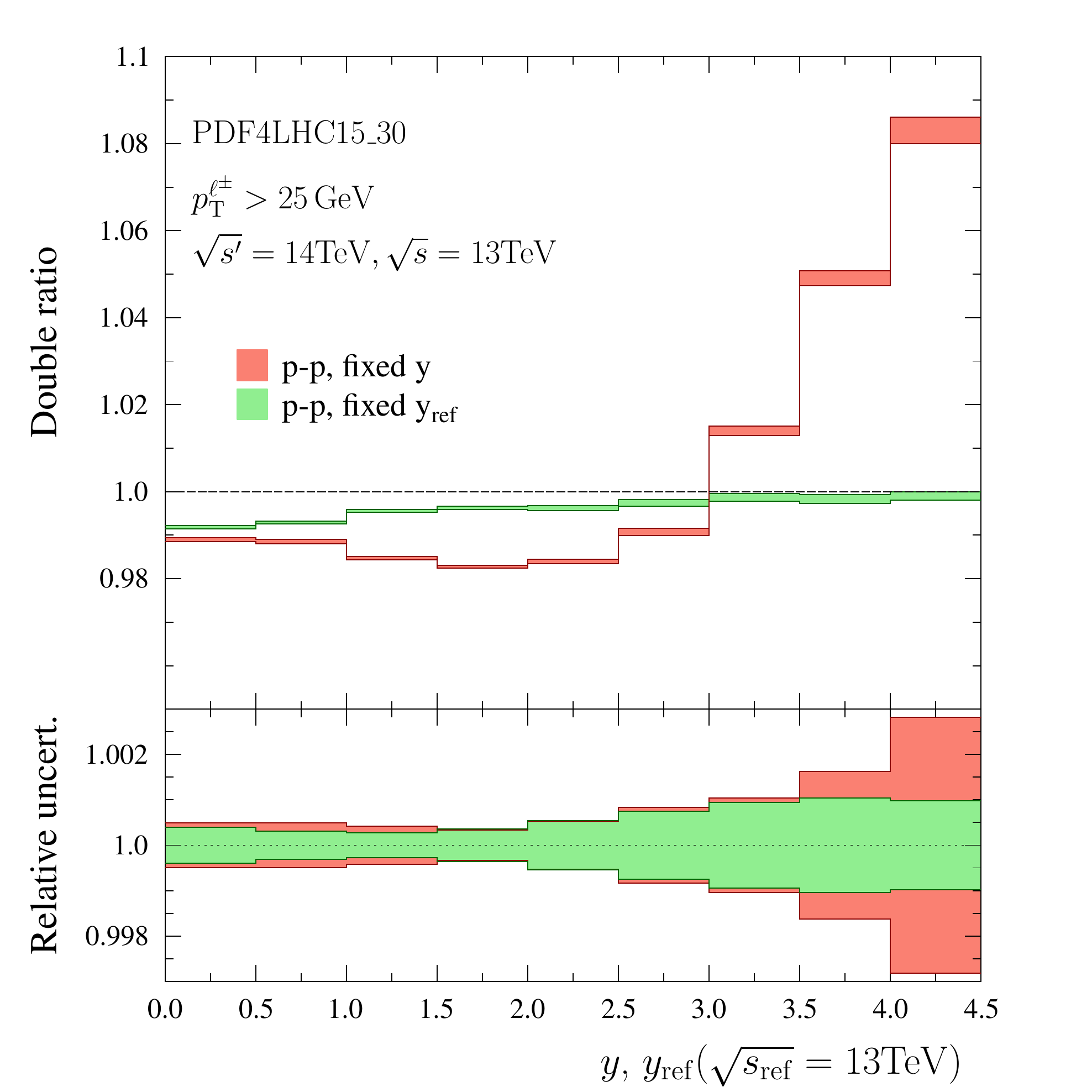}
\caption{As Fig.~\ref{fig:ratio_8_7} but using $\sqrt{s'}=14\,{\rm TeV}$ and $\sqrt{s}=13\,{\rm TeV}$.}
\label{fig:ratio_14_13}
\end{figure*} 

The results are shown in Fig.~\ref{fig:ratio_8_7} ($\sqrt{s'}=8\,{\rm TeV}$, $\sqrt{s}=7\,{\rm TeV}$), Fig.~\ref{fig:ratio_13_8} ($\sqrt{s'}=13\,{\rm TeV}$, $\sqrt{s}=8\,{\rm TeV}$), and Fig.~\ref{fig:ratio_14_13} ($\sqrt{s'}=14\,{\rm TeV}$, $\sqrt{s}=13\,{\rm TeV}$). The histograms in red indicate the outcome when the ratios are taken at fixed rapidity intervals and the green ones correspond to making the ratios at fixed $y_{\rm ref}$ (equivalent to fixed $\xi_1$). One can observe that in the case of ${\rm W}^-$ production and the double ratio the PDF uncertainties indeed tend to cancel out better when the ratios are taken at fixed $y_{\rm ref}$. For ${\rm W}^+$ production it appears that there is no definite advantage (in the sense that PDF uncertainties would decrease) in binning as a function of $y_{\rm ref}$. We attribute this to the fact that in the case of ${\rm W}^+$, the integrand (in Eq.~(\ref{eq1})) in $x$ is broader for ${\rm W}^+$ production than what it is for ${\rm W}^-$ production and the PDF uncertainties do not cancel as effectively.

The LHCb collaboration has recently reported \cite{Aaij:2015zlq} ratios similar to ones discussed here (though integrated over the rapidity interval $2<y<4.5$), and has observed some deviations between the measurements and NLO calculations. Our results suggest that by making the rapidity intervals equal in $y_{\rm ref}$, the PDF uncertainties especially in the double ratio can be suppressed and the significance of the measurement thereby increased.\footnote{An even better precision could be attained by considering the ratios of \emph{total} cross sections \cite{Mangano:2012mh} which, however, are more difficult to measure for the finite acceptance of the experimental apparatuses.}

\section{Summary}
\label{Summary}

We have discussed the scaling properties of inclusive charged leptons from decays of W bosons created in hadronic collisions. Based on the leading-order estimate, we have found that the $\sqrt{s}$ dependence of cross sections in forward/backward directions at fixed value of scaling variable $\xi_{1,2} = (M_{\rm W}/\sqrt{s})e^{\pm y}$ should approximately obey a one-parameter power law, in which the scaling exponent is approximately independent of the lepton charge and reflects the slope of the small-$x$ PDFs. Consequently, the lepton charge asymmetries at different centre-of-mass energies are predicted to be approximately same at fixed $\xi_{1,2}$. Moreover, lepton charge asymmetries in different collision systems are related: at large positive (negative) $y$ the lepton charge asymmetry depends effectively only on the nature of the forward- (backward-) going nucleon or nucleus. A comparison with the experimental data from LHC and Tevatron confirms that the derived scaling laws are indeed able to capture very well the behaviour of the data.

While these scaling laws by no means serve as a replacement for accurate (NLO and beyond) calculations, the possibility of a direct comparison of various data should be useful in e.g. checking the mutual compatibility since by fixing $\xi_1$ ($\xi_2$) one forces the PDFs to be sampled at approximately the same regions of $x_1$ ($x_2$) independently of $\sqrt{s}$. This, as we demonstrated, can in turn be taken advantage of by reducing PDF uncertainties in ratios of cross sections measured at different $\sqrt{s}$. This could increase the sensitivity of the experiments e.g. to possible contributions from physics beyond the Standard Model.

\section*{Acknowledgments}

We would like to thank Rapha\"el Granier de Cassagnac for discussions. We acknowledge CSC (IT Center for Science in Espoo, Finland) for computational resources. The work of \'EC is supported by the European Research Council, under the ``QuarkGluonPlasmaCMS'' \#259612 grant.


\begin{thebibliography}{99}

\bibitem{Mangano:2015ejw}
  M.~L.~Mangano,
  arXiv:1512.00220 [hep-ph].

\bibitem{Berger:1988tu}
  E.~L.~Berger, F.~Halzen, C.~S.~Kim and S.~Willenbrock,
  Phys.\ Rev.\ D {\bf 40} (1989) 83
   [Phys.\ Rev.\ D {\bf 40} (1989) 3789].
  
\bibitem{Martin:1988aj}
  A.~D.~Martin, R.~G.~Roberts and W.~J.~Stirling,
  Mod.\ Phys.\ Lett.\ A {\bf 4} (1989) 1135.

\bibitem{Abe:1998rv}
  F.~Abe {\it et al.}  [CDF Collaboration],
  Phys.\ Rev.\ Lett.\  {\bf 81} (1998) 5754
  [hep-ex/9809001].

\bibitem{Acosta:2005ud}
  D.~Acosta {\it et al.}  [CDF Collaboration],
  Phys.\ Rev.\ D {\bf 71} (2005) 051104
  [hep-ex/0501023].

\bibitem{Abazov:2008qv}
  V.~M.~Abazov {\it et al.}  [D0 Collaboration],
  Phys.\ Rev.\ Lett.\  {\bf 101} (2008) 211801
  [arXiv:0807.3367 [hep-ex]].

\bibitem{Abazov:2007pm}
  V.~M.~Abazov {\it et al.}  [D0 Collaboration],
  Phys.\ Rev.\ D {\bf 77} (2008) 011106
  [arXiv:0709.4254 [hep-ex]].
 
\bibitem{Abazov:2013rja}
  V.~M.~Abazov {\it et al.}  [D0 Collaboration],
  Phys.\ Rev.\ D {\bf 88} (2013) 091102
  [arXiv:1309.2591 [hep-ex]].
  
\bibitem{D0:2014kma}
  V.~M.~Abazov {\it et al.}  [D0 Collaboration],
  Phys.\ Rev.\ D {\bf 91} (2015) 3,  032007
  [arXiv:1412.2862 [hep-ex]].

\bibitem{Aad:2010yt}
  G.~Aad {\it et al.}  [ATLAS Collaboration],
  JHEP {\bf 1012} (2010) 060
  [arXiv:1010.2130 [hep-ex]].
  
\bibitem{Aad:2011yna}
  G.~Aad {\it et al.}  [ATLAS Collaboration],
  Phys.\ Lett.\ B {\bf 701} (2011) 31
  [arXiv:1103.2929 [hep-ex]].

\bibitem{Chatrchyan:2013mza}
  S.~Chatrchyan {\it et al.}  [CMS Collaboration],
  Phys.\ Rev.\ D {\bf 90} (2014) 3,  032004
  [arXiv:1312.6283 [hep-ex]].

\bibitem{Khachatryan:2016pev}
  V.~Khachatryan {\it et al.} [CMS Collaboration],
  arXiv:1603.01803 [hep-ex].
  
\bibitem{Aaij:2014wba}
  R.~Aaij {\it et al.}  [LHCb Collaboration],
  JHEP {\bf 1412} (2014) 079
  [arXiv:1408.4354 [hep-ex]].
  
\bibitem{Aaij:2015zlq}
  R.~Aaij {\it et al.} [LHCb Collaboration],
  JHEP {\bf 1601} (2016) 155
  [arXiv:1511.08039 [hep-ex]].
 
\bibitem{Anastasiou:2003ds}
  C.~Anastasiou, L.~J.~Dixon, K.~Melnikov and F.~Petriello,
  Phys.\ Rev.\ D {\bf 69} (2004) 094008
  [hep-ph/0312266].

\bibitem{Catani:2009sm}
  S.~Catani, L.~Cieri, G.~Ferrera, D.~de Florian and M.~Grazzini,
  Phys.\ Rev.\ Lett.\  {\bf 103} (2009) 082001
  [arXiv:0903.2120 [hep-ph]].

\bibitem{Ball:2010gb}
  R.~D.~Ball {\it et al.}  [NNPDF Collaboration],
  Nucl.\ Phys.\ B {\bf 849} (2011) 112
   [Nucl.\ Phys.\ B {\bf 854} (2012) 926]
   [Nucl.\ Phys.\ B {\bf 855} (2012) 927]
  [arXiv:1012.0836 [hep-ph]].

\bibitem{Lai:2010vv}
  H.~L.~Lai, M.~Guzzi, J.~Huston, Z.~Li, P.~M.~Nadolsky, J.~Pumplin and C.-P.~Yuan,
  Phys.\ Rev.\ D {\bf 82} (2010) 074024
  [arXiv:1007.2241 [hep-ph]].

\bibitem{Khachatryan:2015hha}
  V.~Khachatryan {\it et al.} [CMS Collaboration],
  Phys.\ Lett.\ B {\bf 750} (2015) 565
  doi:10.1016/j.physletb.2015.09.057
  [arXiv:1503.05825 [nucl-ex]].

  
\bibitem{Zhu:2015kpa}
  J.~Zhu [ALICE Collaboration],
  J.\ Phys.\ Conf.\ Ser.\  {\bf 612} (2015) 1,  012009.

\bibitem{ATLAS_W_pPb}
  The ATLAS collaboration,
  ATLAS-CONF-2015-056.
  
\bibitem{Eskola:2009uj}
  K.~J.~Eskola, H.~Paukkunen and C.~A.~Salgado,
  JHEP {\bf 0904} (2009) 065
  [arXiv:0902.4154 [hep-ph]].

\bibitem{Hirai:2007sx}
  M.~Hirai, S.~Kumano and T.-H.~Nagai,
  Phys.\ Rev.\ C {\bf 76} (2007) 065207
  [arXiv:0709.3038 [hep-ph]].

\bibitem{Kovarik:2015cma}
  K.~Kovarik {\it et al.},
  arXiv:1509.00792 [hep-ph].

\bibitem{deFlorian:2011fp}
  D.~de Florian, R.~Sassot, P.~Zurita and M.~Stratmann,
  Phys.\ Rev.\ D {\bf 85} (2012) 074028
  [arXiv:1112.6324 [hep-ph]].
  
\bibitem{Aad:2014bha}
  G.~Aad {\it et al.}  [ATLAS Collaboration],
  Eur.\ Phys.\ J.\ C {\bf 75} (2015) 1,  23
  [arXiv:1408.4674 [hep-ex]].

\bibitem{Chatrchyan:2012nt}
  S.~Chatrchyan {\it et al.}  [CMS Collaboration],
  Phys.\ Lett.\ B {\bf 715} (2012) 66
  [arXiv:1205.6334 [nucl-ex]].

\bibitem{CMS:2012aa}
  S.~Chatrchyan {\it et al.}  [CMS Collaboration],
  Eur.\ Phys.\ J.\ C {\bf 72} (2012) 1945
  [arXiv:1202.2554 [nucl-ex]].

\bibitem{Aad:2015wga}
  G.~Aad {\it et al.} [ATLAS Collaboration],
  JHEP {\bf 1509} (2015) 050
  doi:10.1007/JHEP09(2015)050
  [arXiv:1504.04337 [hep-ex]].
  
\bibitem{Abelev:2014laa}
  B.~B.~Abelev {\it et al.}  [ALICE Collaboration],
  Phys.\ Lett.\ B {\bf 736} (2014) 196
  [arXiv:1401.1250 [nucl-ex]].

\bibitem{CMS:2012rba}
  CMS Collaboration [CMS Collaboration],
  CMS-PAS-HIN-12-004.

\bibitem{Aad:2014bxa}
  G.~Aad {\it et al.}  [ATLAS Collaboration],
  Phys.\ Rev.\ Lett.\  {\bf 114} (2015) 7,  072302
  [arXiv:1411.2357 [hep-ex]].
  
\bibitem{Adam:2015ewa}
  J.~Adam {\it et al.}  [ALICE Collaboration],
  Phys.\ Lett.\ B {\bf 746} (2015) 1
  [arXiv:1502.01689 [nucl-ex]].
  
\bibitem{Paukkunen:2010qg}
  H.~Paukkunen and C.~A.~Salgado,
  JHEP {\bf 1103} (2011) 071
  [arXiv:1010.5392 [hep-ph]].

  
\bibitem{Ru:2014yma}
  P.~Ru, B.~W.~Zhang, L.~Cheng, E.~Wang and W.~N.~Zhang,
  J.\ Phys.\ G {\bf 42} (2015) no.8,  085104
  doi:10.1088/0954-3899/42/8/085104
  [arXiv:1412.2930 [nucl-th]].
  
\bibitem{Ru:2015pfa}
  P.~Ru, B.~W.~Zhang, E.~Wang and W.~N.~Zhang,
  Eur.\ Phys.\ J.\ C {\bf 75} (2015) no.9,  426
  doi:10.1140/epjc/s10052-015-3652-x
  [arXiv:1505.08106 [nucl-th]].
  
\bibitem{Aurenche:1980tp}
  P.~Aurenche and J.~Lindfors,
  Nucl.\ Phys.\ B {\bf 185} (1981) 274.

\bibitem{Baer:1990qy}
  H.~Baer and M.~H.~Reno,
  Phys.\ Rev.\ D {\bf 43} (1991) 2892.

\bibitem{Gluck:1998xa}
  M.~Gluck, E.~Reya and A.~Vogt,
  Eur.\ Phys.\ J.\ C {\bf 5} (1998) 461
  [hep-ph/9806404].

  
\bibitem{Ball:1994du}
  R.~D.~Ball and S.~Forte,
  Phys.\ Lett.\ B {\bf 335} (1994) 77
  doi:10.1016/0370-2693(94)91561-X
  [hep-ph/9405320].
  
\bibitem{Gribov:1981ac}	
L.V. Gribov, E.M. Levin, M.G. Ryskin, Nucl.\ Phys.\ B {\bf 188} (1981) 555.

\bibitem{Dokshitzer:1977sg}
  Y.~L.~Dokshitzer,
  Sov.\ Phys.\ JETP {\bf 46} (1977) 641
   [Zh.\ Eksp.\ Teor.\ Fiz.\  {\bf 73} (1977) 1216].

\bibitem{Gribov:1972ri}
  V.~N.~Gribov and L.~N.~Lipatov,
  Sov.\ J.\ Nucl.\ Phys.\  {\bf 15} (1972) 438
   [Yad.\ Fiz.\  {\bf 15} (1972) 781].

\bibitem{Gribov:1972rt}
  V.~N.~Gribov and L.~N.~Lipatov,
  Sov.\ J.\ Nucl.\ Phys.\  {\bf 15} (1972) 675
   [Yad.\ Fiz.\  {\bf 15} (1972) 1218].

\bibitem{Altarelli:1977zs}
  G.~Altarelli and G.~Parisi,
  Nucl.\ Phys.\ B {\bf 126} (1977) 298.

\bibitem{Adloff:2001rw}
  C.~Adloff {\it et al.} [H1 Collaboration],
  Phys.\ Lett.\ B {\bf 520} (2001) 183
  [hep-ex/0108035].
  
\cite{Campbell:2010ff}
\bibitem{Campbell:2010ff}
  J.~M.~Campbell and R.~K.~Ellis,
  Nucl.\ Phys.\ Proc.\ Suppl.\  {\bf 205-206} (2010) 10
  [arXiv:1007.3492 [hep-ph]].

\bibitem{Aad:2011dm}
  G.~Aad {\it et al.}  [ATLAS Collaboration],
  Phys.\ Rev.\ D {\bf 85} (2012) 072004
  [arXiv:1109.5141 [hep-ex]].

\bibitem{Butterworth:2015oua}
  J.~Butterworth {\it et al.},
  J.\ Phys.\ G {\bf 43} (2016) 023001
  doi:10.1088/0954-3899/43/2/023001
  [arXiv:1510.03865 [hep-ph]].

\bibitem{Buckley:2014ana}
  A.~Buckley, J.~Ferrando, S.~Lloyd, K.~Nordström, B.~Page, M.~Rüfenacht, M.~Schönherr and G.~Watt,
  Eur.\ Phys.\ J.\ C {\bf 75} (2015) 132
  doi:10.1140/epjc/s10052-015-3318-8
  [arXiv:1412.7420 [hep-ph]].
  
\bibitem{Gao:2013bia}
  J.~Gao and P.~Nadolsky,
  JHEP {\bf 1407} (2014) 035
  doi:10.1007/JHEP07(2014)035
  [arXiv:1401.0013 [hep-ph]].

\bibitem{Dulat:2015mca}
  S.~Dulat {\it et al.},
  Phys.\ Rev.\ D {\bf 93} (2016) 033006
  [arXiv:1506.07443 [hep-ph]].

\bibitem{Harland-Lang:2014zoa}
  L.~A.~Harland-Lang, A.~D.~Martin, P.~Motylinski and R.~S.~Thorne,
  Eur.\ Phys.\ J.\ C {\bf 75} (2015) 5,  204
  [arXiv:1412.3989 [hep-ph]].

\bibitem{Ball:2014uwa}
  R.~D.~Ball {\it et al.} [NNPDF Collaboration],
  JHEP {\bf 1504} (2015) 040
  doi:10.1007/JHEP04(2015)040
  [arXiv:1410.8849 [hep-ph]].
  
\bibitem{Mangano:2012mh}
  M.~L.~Mangano and J.~Rojo,
  JHEP {\bf 1208} (2012) 010
  [arXiv:1206.3557 [hep-ph]].

  
\end{thebibliography}
\end{document}